\documentclass[twocolumn,epjc3]{svjour3}
\usepackage{graphics}
\usepackage{multicol}
\usepackage{cuted}
\usepackage{flushend}

\usepackage{amsfonts,amsmath,amssymb}

\usepackage{microtype}
\usepackage{bbm}
\usepackage{amsmath}
\usepackage{mathtools}
\usepackage{caption}
\usepackage{subcaption}
\usepackage{amssymb}
\usepackage{mathrsfs}       
\usepackage[pdftex]{graphicx}
\usepackage{pgfplots}
\pgfplotsset{width=10cm,compat=1.9}
\usepackage{xargs}       
\usepackage{wasysym}
\usepackage{engrec}
\usepackage{enumerate}
\usepackage{appendix}
\usepackage{empheq}
\usepackage{float}
\usepackage{stmaryrd}
\usepackage[parfill]{parskip}
\usepackage[breaklinks,hidelinks]{hyperref}
\usepackage{titletoc}
\usepackage{makecell}
\usepackage{lscape}
\usepackage{tensor}
\usepackage{blkarray}
\usepackage{multirow}
\usepackage{bigstrut}
\usetikzlibrary{plotmarks}
\usepackage{comment}
\usepackage[version=4]{mhchem}
\usepackage[ruled]{algorithm2e}
\usepackage{soul}
\usepackage{listings}
\usepackage{cancel}

\SetKwData{Input}{}

\SetKwHangingKw{KwSolve}{Solve} 
\SetKwHangingKw{KwReturn}{Return} 
\SetKwHangingKw{KwBreak}{break} 
    \SetKwRepeat{KwRepeat}{do}{while}
    \SetKwFor{KwFor}{for}{do}{end for}

\newcommand{\nucl}[2]{\({}^{#2}\mathrm{#1}\)}

\newcommand{\clebsch}[6]{C_{#1 #2 #3 #4}^{#5 #6}}

\newenvironment{customlegend}[1][]{%
    \begingroup
    \csname pgfplots@init@cleared@structures\endcsname
    \pgfplotsset{#1}%
}{%
    \csname pgfplots@createlegend\endcsname
    \endgroup
}%
\def\addlegendimage{\csname pgfplots@addlegendimage\endcsname}

\newcommand{\correction}[2]{{\color{black}#2}}
\makeatletter
\newcommand*{\transpose}{{\mathpalette\@transpose{}}}
\newcommand*{\@transpose}[2]{\raisebox{\depth}{$\m@th#1\intercal$}}
\makeatother

\begin{document}
\title{Tensor factorization in \textit{ab initio} many-body calculations}
\subtitle{Triaxially-deformed (B)MBPT calculations in large bases}

\author{
}                 

\thankstext{em:mf}{\email{mikael.frosini@cea.fr}}
\thankstext{em:td}{\email{thomas.duguet@cea.fr}}
\thankstext{em:pt}{\email{pierre.tamagno@cea.fr}}

\author{
M. Frosini\thanksref{ad:des}
\and T. Duguet\thanksref{ad:saclay,ad:kul}
\and P. Tamagno\thanksref{ad:des}
}

\institute{
\label{ad:des}
CEA, DES, IRESNE, DER, SPRC, LEPh, France
\and
\label{ad:saclay}
IRFU, CEA, Universit\'e Paris-Saclay, 91191 Gif-sur-Yvette, France 
\and
\label{ad:kul}
KU Leuven, Department of Physics and Astronomy, Instituut voor Kern- en Stralingsfysica, 3001 Leuven, Belgium 
}

\date{Received: \today{} / Revised version: date}

\maketitle
%
%
\begin{abstract}
Whether for fundamental studies or nuclear data evaluations, first-principle calculations of atomic nuclei constitute the path forward. Today, performing \textit{ab initio} calculations (a) of heavy nuclei, (b) of doubly open-shell nuclei or (c) with a sub-percent accuracy  is at the forefront of nuclear structure theory. While combining any two of these features constitutes a major challenge, addressing the three at the same time is currently impossible. From a numerical standpoint, these challenges relate to the necessity to handle (i) very large single-particle bases and (ii) mode-6, \textit{i.e.} three-body, tensors (iii) that must be stored repeatedly. Performing second-order many-body perturbation theory(ies) calculations based on triaxially deformed and superfluid reference states of doubly open-shell nuclei up to mass $A=72$, the present work achieves a significant step forward by addressing challenge (i). To do so, the memory and computational cost associated with the handling of large tensors is scaled down via the use of tensor factorization techniques. The presently used factorization format is based on a randomized singular value decomposition that does not require the computation and storage of the very large initial tensor. The procedure delivers an inexpensive and controllable approximation to the original problem, as presently illustrated for calculations that could not be performed without tensor factorization. With the presently developed technology at hand, one can envision to perform calculations of yet heavier doubly open-shell nuclei at sub-percent accuracy in a foreseeable future.
\end{abstract}

\section{Introduction}

The \textit{ab initio} description of the atomic nucleus is rapidly extending its reach over the nuclear chart, mainly thanks to the development and implementation of so-called expansion many-body methods~\cite{Hergert:2020bxy,Dickhoff2004,Hergert16a,Hagen2010,tichai2020many}. Such methods reformulate the $A$-body Schrödinger equation within the second quantization framework in terms of tensors, \textit{i.e.} multi-dimensional arrays, and tensor networks. These tensors belong to two categories. First are the {\it known} tensors carrying the information of inter-nucleon interactions that constitute the input to the problem. Second are  the {\it unknown} many-body tensors parametrizing the $A$-body state and constituting the output of the problem. Eventually, observables are computed as tensor networks combining both sets of tensors. 

While the maximum mode, \textit{i.e.}  number of indices, $2k_{\text{max}}$ among the input tensors relates to the interaction of highest $k_{\text{max}}$-body character, the maximum mode $2q_{\text{max}}$ of the output tensors relates to the truncation order, \textit{i.e.} the target accuracy, chosen to compute the $A$-body state of interest.

The indices of the input tensors typically run over an appropriate basis of the one-body Hilbert space \(\mathcal H_1\) whose dimension, in principle infinite, is made finite in any actual numerical application. Of course, this finite dimension $N$ must be taken large enough for the resulting error to remain below a chosen threshold. Typically, the needed value of $N$ increases with the mass of the system. Effectively, the ``naive'' dimension $N$ can be reduced to a lower dimension $\tilde{N}$ by choosing a basis whose symmetry properties allow one to best exploit the physical symmetries of the Hamiltonian\footnote{More specifically, the input tensors expressed in that basis display a block-diagonal structure, \textit{i.e.} an explicit sparsity, such that each block is characterized by the smaller effective dimension $\tilde{N}$. See Sec.~\ref{discussion_dim} for details.}. 

Writing out the tensor networks necessary to determine the output tensors and to compute observables requires to go from the initial basis, where input tensors are provided, to a quasi-particle (qp) basis in which those tensor networks are formulated. While the quasi-particle basis shares the same ``naive'' dimension $N$, its degree of symmetry may be lower in the case where the expansion method at play exploits the concept of spontaneous symmetry breaking to capture strong static correlations in open-shell nuclei~\cite{Soma:2013xha,Tichai18BMBPT,Yao:2019rck,Novario:2020kuf,Hagen22a,Frosini22c,Tichai:2023epe}. Thus, the effective dimension $\tilde{N}_{\text{qp}}$ at play in the qp basis ranges from $\tilde{N}$ back to $N$ depending on the specific many-body framework used.

Eventually, both sets of tensors expressed in the quasi-particle basis must in principle be computed, stored and combined to solve the $A$-body Schrödinger equation to a given accuracy. The complexity, \textit{i.e.} memory and CPU costs, associated with these three tasks depends strongly on all of the above mentioned parameters, \textit{i.e.}
\begin{enumerate}
\item The maximum modes of the tensors
\begin{enumerate}
\item $2k_{\text{max}}$ for input tensors,
\item $2q_{\text{max}}$ for output tensors.
\end{enumerate}
\item The effective dimensions of the bases 
\begin{enumerate}
\item $\tilde{N}$ for input tensors,
\item $\tilde{N}_{\text{qp}}$ for both input and output tensors.
\end{enumerate}
\end{enumerate}
Depending on the (combined) values of these parameters, the tasks of computing, storing and combining the tensors in realistic calculations range from being trivial to being undoable. Consequently, performing \textit{ab initio} calculations of (a) heavy nuclei, (b) doubly open-shell nuclei or (c) with a sub-percent accuracy  is at the forefront of \textit{ab initio} nuclear structure theory today, whereas combining any two of these features constitutes a major challenge. Combining the three of them is currently impossible. 

Lowering the memory and CPU footprints of nuclear \textit{ab initio} calculations is thus of great present interest. Several avenues are currently being pursued~\cite{Ro09,Tichai:2018qge,Tichai:2019ksh,Hoppe:2020elo,Porro:2021rjp,Hoppe:2021xqv} to eventually achieve this goal. The one followed in the present work relates to the use of (truncated) tensor factorization ((T)TF) techniques~\cite{Tichai:2018eem,Tichai:2019ksh,Tichai:2021rtv,Tichai:2022mqn,Tichai:2023dda}. Several steps in this direction have already been taken recently. In Ref.~\cite{Tichai:2018eem}, the TF of the many-body (\textit{i.e.} two-body) tensor at play in second-order spherical many-body perturbation theory ($_\text{s}$MBPT(2)) calculations of doubly closed-shell nuclei was investigated using a (unrealistically) small effective basis dimension ($\tilde{N}_{\text{qp}}=30$ originating from $N=140$). In Ref.~\cite{Tichai:2019ksh}, the same scheme was extended to second-order spherical Bogoliubov MBPT ($_\text{s}$BMBPT(2)) calculations of singly open-shell nuclei. In Ref.~\cite{Zare:2022cdw}, a more versatile factorization technique was employed to push $_\text{s}$MBPT(2) calculations of doubly closed-shell nuclei in larger bases (up to $\tilde{N}_{\text{qp}}=216$ originating from $N=2720$) allowing calculation up to $^{132}$Sn. Based on such promising efforts, the goal is now to develop TF amenable to much larger basis dimensions, \textit{i.e.} $\tilde{N}_{\text{qp}}\approx N \in [1000,4000]$, which are necessary to perform realistic calculations of mid-mass doubly open-shell nuclei. The sizes of the tensors to handle are obviously orders-of-magnitude more challenging. 

The above calculations were performed  with rank-reduced three-nucleon (3N) interactions (effectively reducing from $k_{\text{max}}=3$ to $k_{\text{max}}=2$) at (B)MBPT(2) level ($q_{\text{max}}=2$). The real challenge is eventually to perform high-accuracy calculations ($q_{\text{max}}=3$), possibly with genuine 3N ($k_{\text{max}}=3$) or even four-nucleon (4N) ($k_{\text{max}}=4$) interactions, which is greatly more challenging. Recently, another versatile TF format was put forward and applied to 3N interaction matrix elements ($k_{\text{max}}=3$) expressed in the momentum basis~\cite{Tichai:2023dda}. The error originating from TTF 3N matrix elements  was shown to lead to small errors in the realistic computation ($\tilde{N}_{\text{qp}}= 240$ originating from $N=2720$)\footnote{Note that, for given a value of $\tilde{N}$, not all elements of the 3N interaction tensor are considered to begin with in state-of-the-art \textit{ab initio} calculations. Indeed, an initial truncation of the tensor is realized based on the range covered by its first, resp. last, three indices~\cite{Hergert:2020bxy}. See Sec.~\ref{rankred} for details.}  of ground-state energies and charge radii of doubly-closed-shell nuclei up to $^{132}$Sn. Calculations were performed using MBPT(2) as well as non-perturbative in-medium similarity renormalization group (IMSRG) at the two-body, \textit{i.e.} IMSRG(2), truncation level ($q_{\text{max}}=2$). However, the compression of the input 3N interaction tensor could not actually be exploited in the many-body calculation itself given that the factorization performed in the momentum basis does not propagate to the qp bases relevant to finite-nuclei calculations.  

The present work aims at pushing TF techniques in several combined directions. The first objective is to adapt the TF format put forward in Ref.~\cite{Tichai:2023dda} to compute many-body tensors directly in the qp basis relevant to finite-nuclei calculations. The goal is to achieve such a task to study doubly open-shell nuclei based on large bases while (potentially) breaking both U(1) and  SU(2) symmetry associated with particle number and angular momentum conservation, respectively. In particular, both $J^2$ {\it and} $J_z$ are allowed to break, \textit{i.e.} explicitly triaxial calculations up to $^{72}$Kr are considered. In this context, no gain can be obtained by reducing the effective range of qp indices based on symmetry considerations, \textit{i.e.} one has to face the full initial basis dimension $\tilde{N}_{\text{qp}}=N \in [2000,4000]$. Consequently, even a two-body (\textit{i.e.} mode-4) tensor is eventually too large to be computed and stored in the qp basis. Thus, one key objective of the present work is to perform the factorization of the many-body tensors at play without ever computing and storing them explicitly.

The above program is presently implemented based on (B)MBPT(2) calculations \correction{\footnote{The four specific settings under consideration are listed in Tab.~\ref{tab:methods}.}}{(see. Tab.~\ref{tab:methods} for a detailed account of the four settings under consideration)} and thus strongly benefits from two simplifications. First, no three-body (\textit{i.e.} mode-6) tensor arises in second-order calculations, \textit{i.e.} $q_{\text{max}}=2$. This avoids for now to accumulate the challenges of going to $q_{\text{max}}=3$ and to large $\tilde{N}_{\text{qp}}=N$ at the same time. Second, the many-body (\textit{i.e.} two-body) tensor is a known analytical functional of the input tensors in perturbation theory. This bypasses the need to solve any dynamical equation to compute the two-body tensor, which is the source of the larger (memory and CPU) cost of non-perturbative methods for a given $q_{\text{max}}$. The further increase of complexity due to going to (B)MBPT(3) and to non-perturbative, \textit{e.g.} (B)CC \correction{}{((Bogoliubov) Coupled Cluster)~\cite{signoracci2015ab,Tichai:2023epe}}, methods is left to future studies\correction{}{\footnote{Some factorized formalisms have been recently derived in the context of quantum chemistry~\cite{Hohenstein_2022}, but are still yet at a preliminary stage.}}. In the latter case, benefiting from TF will require to apply it to the dynamical equations themselves in order to solve directly for the {\it factors} of the {\it a priori} factorized unknown many-body tensors.

\begin{table}
    \centering
\begin{tabular}{c|c| c |c}
  \hline
    \multicolumn{2}{c }{} & \multicolumn{2}{|c }{Symmetry}  \\
     \hline
  Acronym & Reference state $| \Phi \rangle$  & Gauge & Rotational   \\
    \hline
    \hline
    $_\text{a}$MBPT &  Slater determinant  & Yes ($A$) & Axial ($M=0$) \\
    \hline
    $_\text{t}$MBPT &  Slater determinant  & Yes ($A$) & None ($\cancel{M}$) \\
    \hline
    $_\text{a}$BMBPT &  Bogoliubov state  & No ($\cancel{A}$) & Axial ($M=0$)  \\
    \hline
    $_\text{t}$BMBPT &  Bogoliubov state  & No ($\cancel{A}$) & None ($\cancel{M}$) \\
    \hline
    \end{tabular}
    \captionof{table}{Many-body methods presently employed.}
    \label{tab:methods}
\end{table}

The present paper is organized as follows. Section~\ref{secindices} details the characteristics of the bases at play in the problem. Section~\ref{secHtensors} discusses the characteristics of the Hamiltonian tensors and the various techniques already in use to reduce their naive complexity prior to starting the actual many-body calculation. While Sec.~\ref{sectMBtensors} explains briefly the appearance of many-body tensors in expansion many-body methods, Sec.~\ref{formalBMBPT} particularizes the discussion to axial $_\text{a}$(B)MBPT(2) and triaxial $_\text{t}$(B)MBPT(2) calculations of present interest. Having characterized the challenges at play, Sec.~\ref{SecTF} introduces the specific tensor factorization scheme used to overcome them. Next, Sec.~\ref{sec:results} presents the numerical results illustrating the benefits of this (truncated) TF format along with its promises for even more challenging settings in the future. While conclusions of the present work are provided in Sec.~\ref{secconclusions}, an appendix details important technical aspects of the tensor factorization.

\section{Tensor indices}
\label{secindices}

\begin{table}[t!]
\def\arraystretch{1.6}
\centering
\begin{tabular}{c | c | c }
\hline \hline 
$e_\text{max}$ & $N$ & $_\text{s}\tilde{N}$    \\
\hline \hline
2 & 40 & 12  \\
4 & 140 & 30   \\
6 & 336 & 56    \\
8 & 660 & 90      \\
10 & 1144 & 132  \\
12 & 1820 & 182  \\
14 & 2720 & 240      \\
16 & 3876 & 306  \\
\hline \hline
\end{tabular}
\caption{Dimension $N$ of a spherical basis driven by the truncation parameter $e_\text{max}$ and the corresponding effective dimension $_\text{s}\tilde{N}$ at play whenever spherical symmetry can be exploited via angular momentum coupling techniques; see Sec.~\ref{AMCSec} for details.}
\label{tabledimensions}
\end{table}

One must first characterize the different sets of indices labeling the tensors at play in the present work. 

\subsection{Rotational symmetry}
\label{rot}

One first typically considers a spherical basis whose states are characterized by the set of quantum numbers 
\begin{align}
\alpha \equiv (n_\alpha, \ell_\alpha, j_\alpha, m_\alpha, t_\alpha) \, , \label{set1}
\end{align}
where $n_\alpha$ denotes a principal quantum number, $\ell_\alpha$ the orbital angular-momentum quantum number, $j_\alpha$ the total angular-momentum quantum number and $m_\alpha\in[-j_\alpha,j_\alpha]$ its projection on the $z$ axis, as well as $t_\alpha$ the isospin projection. The dimension $N$, \textit{i.e.} the range of the indices, is set by selecting the states according to $0 \leq e_\alpha \leq e_\text{max}$ with $e_\alpha \equiv 2n_\alpha +\ell_\alpha$. The values of $N$ corresponding to $2\leq e_\text{max} \leq 16$ are displayed in Tab.~\ref{tabledimensions}.

With such a basis at hands, one further considers the set of indices defined through the subset of quantum numbers
\begin{align}
\tilde \alpha \equiv (n_\alpha, \ell_\alpha, j_\alpha, t_\alpha) \, , \label{set1'}
\end{align}
differing from $\alpha$ by the removal of the magnetic quantum number $m_\alpha$. As a result, the reduced set runs over an effective dimension $_\text{s}\tilde{N}$ that is much smaller than $N$. The corresponding values are also provided in Tab.~\ref{tabledimensions}. 

\subsection{Breaking rotational symmetry}
\label{rotno}

Indices characteristic of lower spatial symmetries are eventually necessary. Whenever breaking rotational symmetry in three dimensions while maintaining it around the $z$ axis, the indices of interest become
\begin{align}
k \equiv (n'_k, \pi_k, m_k, t_k) \, , \label{set2}
\end{align}
where $n'_k$ denotes a novel principal quantum number and $\pi_k$ the parity quantum number. While $m_k$ remains a good quantum number, this is not anymore the case for $j_k$ and $\ell_k$.

Further breaking axial symmetry leads us to replacing the previous set by
\begin{align}
k \equiv (n''_k, \pi_k, t_k) \, , \label{set3}
\end{align}
such that $m_k$ is not a good quantum number anymore. The last two sets being typically obtained via a unitary transformation of the original set introduced in Eq.~\eqref{set1}, their cardinality is also equal to $N$.

\section{Hamiltonian tensors}
\label{secHtensors}

\subsection{Definition}

\textit{Ab initio} nuclear structure theory aims at finding the eigenstates of the nuclear Hamiltonian
\begin{equation}
    H \equiv T + V + W + \dots \, , \label{H}
\end{equation}
for a system of \(A\) interacting nucleons, where \(T\) denotes the one-body kinetic operator, \(V\) the two-nucleon interaction\footnote{In practice, while $T$ also incorporates the one-body part of the subtracted center-of-mass kinetic energy, $V$ contains its two-body contribution.}, and \(W\) the three-body interaction. 
While not explicitly specified in Eq.~\eqref{H}, \(H\) contains in principle up to \(A\)-nucleon (AN) interaction terms. 

In order to explicitly represent the input Hamiltonian, a (truncated) basis of the one-body Hilbert-space \(\mathcal H_1\) must be specified. In the present work, the eigenbasis of the spherical one-body harmonic oscillator (sHO) Hamiltonian characterized by the frequency $\hbar\omega$, and belonging to the category introduced in Eq.~\eqref{set1}, is employed. Introducing the corresponding set of particle creation and annihilation operators \(\{c,c^\dag\}\), \(T,\ V\) and \(W\) are respectively parameterized by mode-$2$,  mode-$4$ and mode-$6$ tensors according to
\begin{subequations}
    \begin{align}
        T &\equiv \frac 1{(1!)^2} \sum_{\alpha\beta} t_{\alpha\beta} \, c^\dag_{\alpha} c_{\beta} \, ,\\
        V &\equiv \frac 1{(2!)^2} \sum_{\alpha\beta\gamma\delta} v_{\alpha\beta\gamma\delta} \, c^\dag_{\alpha} c^\dag_{\beta} c_{\delta} c_{\gamma} \, ,\\
        W &\equiv \frac 1{(3!)^2} \sum_{\alpha\beta\gamma\delta\zeta\epsilon} w_{\alpha\beta\gamma\delta\zeta\epsilon} \, c^\dag_{\alpha} c^\dag_{\beta}c^\dag_{\gamma} c_{\epsilon}c_{\zeta}c_{\delta} \, ,
    \end{align}
    \label{eq:HamiltonianTensors}
\end{subequations}
where $v_{\alpha\beta\gamma\delta}$ ($w_{\alpha\beta\gamma\delta\zeta\epsilon}$) is anti-symmetric under the exchange of the first or last two (any pair among the first or last three) indices. Obviously, the maximum mode at play in the Hamiltonian defined through Eq.~\eqref{eq:HamiltonianTensors} is set by the 3N interaction, \textit{i.e.} $k_{\text{max}}=3$.

\subsection{Angular momentum coupling}
\label{AMCSec}

The size of mode-$2k$ tensors naively scales with $N$ as $N^{2k}$, which quickly leads to unbearable sizes for the 3N interaction (\textit{i.e.} a mode-6 tensor). The Hamiltonian being rotationally invariant, the mode-$2k$ tensors representing it in the sHO basis are however block-diagonal with respect to the $k$-body total angular momentum $J$. Such an explicit sparsity  can be exploited \textit{a priori} via the application of so-called angular-momentum coupling (AMC) techniques. Taking the simple example of the two-body interaction tensor, the J-coupled tensor is obtained from the initial one by performing the AMC according to\footnote{The AMC of mode-$n$ tensors and of elaborate tensor networks involving many such tensors can be automatized~\cite{Tichai:2020jpk}.}
\begin{align}
  \tilde v^{J}_{\tilde \alpha \tilde \beta \tilde \gamma \tilde \delta} &= \sum_{ \substack{m_{\alpha} m_{\beta} \\ m_{\gamma} m_{\delta} }}  v_{\alpha\beta\gamma\delta} \,   \clebsch{j_{\alpha}}{m_{\alpha}}{j_{\beta}}{m_{\beta}}{J}{M} \clebsch{j_{\gamma}}{m_{\gamma}}{j_{\delta}}{m_{\delta}}{J}{M} \, ,
 \label{eq:twobodyJ}
\end{align}
which is block diagonal with respect to the two-body total angular momentum $J$ and happens to be independent of its projection $M$. Effectively, as a result of the AMC procedure the indices of $\tilde v^{J}_{\tilde \alpha \tilde \beta \tilde \gamma \tilde \delta}$ belong to the {\it reduced} set introduced in Eq.~\eqref{set1'}, which diminishes tremendously their range $_\text{s}\tilde{N}$ and thus the size of the J-coupled tensors. 

While the AMC does not change the mode of the tensor, it reduces tremendously the range of its indices and thus leads to a huge compression of the tensor
. Eventually, the number of expected non-zero elements of a mode-$2k$ tensor is $_\text{s}\tilde{N}^{2k}$, times the number of possible $J$ values, rather than $N^{2k}$. Given the values of $N$ and $_\text{s}\tilde{N}$ given in Tab.~\ref{tabledimensions}, the compression factor increases rapidly with $N$ and $k$, \textit{e.g.} the size of the two-body interaction tensor is reduced by 6 orders of magnitude for $N=1820$ ($e_\text{max}=12$) as can be inferred from the left panel of Fig.~\ref{fig:scaling}.

\subsection{Rank reduction}
\label{rankred}

Dealing with the 3N interaction tensor $W$ in actual many-body computations is already problematic in rather light nuclei even when relying on J-coupled tensors. 

Handling the full J-coupled 3N tensor is beyond reach already for $e_\text{max} \geq 10$. From the outset, a compression is thus performed that consists of ignoring all tensor entries for which the parameter $e_{\alpha\beta\gamma} \equiv e_{\alpha}+e_{\beta}+e_{\gamma}$ of the first (resp. last) three indices is larger than a chosen value $e_\text{3max} \leq 3 e_\text{max}$. While employing $e_\text{3max}$ around 14 or 16 is sufficient to tackle nuclei with $A\leq 80$, moving towards the $A\sim [150,210]$ regime typically requires $e_\text{3max}\sim[24,28]$; see Ref.~\cite{Miyagi2021,Tichai:2023epe} for details.

In addition to this basic IT \correction{}{(Importance Truncation)}-like compression~\cite{Ro09,Tichai:2019ksh,Porro:2021rjp,Hoppe:2021xqv} effectively reducing the range of the mode-6 tensor's indices, it is necessary to lower the mode of the tensor in order to take it into account when computing tensor networks delivering the unknown many-body tensors and observables. In order to do so, it has become customary to proceed to a rank-reduction of the 3N operator via either a (truncated) normal ordering procedure~\cite{RoBi12,Gebrerufael:2015yig}  or via a more general rank-reduction method~\cite{Frosini21a}. Following the latter, it leads to approximating Eq.~\eqref{H} by
\begin{align}
    H \approx& h^{(0)}[\rho] \nonumber  \\
    &+\frac{1}{(1!)^2} \sum_{\alpha\beta} h^{(1)}_{\alpha\beta}[\rho] \, c^\dagger_{\alpha} c_{\beta} \nonumber \\
    &+\frac{1}{(2!)^2} \sum_{\alpha\beta\gamma\delta} h^{(2)}_{\alpha\beta\gamma\delta}[\rho] \, c^\dagger_{\alpha} c^\dagger_{\beta}c_{\delta} c_{\gamma} \, , \label{finalapproxH}
\end{align}
where the mode-0 tensor $h^{(0)}[\rho]$, the mode-2 tensor $h^{(1)}_{\alpha\beta}[\rho]$ and the mode-4 tensor $h^{(2)}_{\alpha\beta\gamma\delta}[\rho]$ are functionals of the initial input tensors and of a symmetry-invariant one-body density matrix $\rho$ associated with an appropriate many-body state~\cite{Frosini21a}. \correction{}{In this work, and in agreement with the conclusions of Ref.~\cite{Frosini21a}, $\rho$ originates from a spherical HFB calculation.} This rank-reduction procedure, which  effectively acts as a basic TF approximating the 3N interaction tensor via mode-$2k$ tensors with $k<3$, typically induces a few percent error in mid-mass nuclei~\cite{Frosini21a}.

While these two basic compression methods have made possible to tremendously advance \textit{ab initio} calculations in recent years, it will eventually be necessary to improve upon them in the future, \textit{e.g.}  replace the rank-reduction method by a more advanced TF of the 3N interaction tensor\footnote{The idea of looking for compact representations of interaction matrix elements is not new. For example, strictly separable 2N interactions~\cite{haftel70a,Duguet:2003yi} or specific representations such as Gaussian functions authorizing analytical factorizations~\cite{Robledo:2010ef} have been used for a long time.}. In this work, however, the handling of the input Hamiltonian tensors is limited to the above, already standard, compression methods such that the highest mode effectively at play for the subsequent many-body calculation is $k_{\text{max}}=2$ in agreement with Eq.~\eqref{finalapproxH}.

\section{Many-body tensors}
\label{sectMBtensors}

The goal is to solve $A$-body Schrödinger's equation 
\begin{equation}
H | \Psi^{\text{A}}_{\mu} \rangle = E^{\text{A}}_{\mu}  | \Psi^{\text{A}}_{\mu} \rangle \, ,
\end{equation}
via an expansion method of choice. In the present work, the target is the nuclear ground-state $| \Psi^{\text{A}}_{0} \rangle$. 

\subsection{Unperturbed state and quasi-particle basis}

A given expansion method is first characterized by the choice of an appropriate partitioning of the Hamiltonian, thus defining the unperturbed part whose exact eigenstates can be determined, its ground-state being the so-called {\it unperturbed state} $|\Phi\rangle$. The present work focuses on so-called {\it single-reference} methods that are based on unperturbed Hamiltonians whose eigenstates belong to the class of {\it product} states\footnote{So-called {\it multi-reference} methods are based on more general classes of unperturbed states and can be formulated while having explicit access to a single eigenstate of the unperturbed Hamiltonian~\cite{Frosini22a}.}. Being interested in the most general set of product states, it is necessary to work with the grand potential \(\Omega\equiv H-\lambda A\) rather than with the Hamiltonian itself, where \(\lambda\) is the chemical potential and \(A\) the particle number operator. Given that $H$ and $A$ commute, the Hamiltonian and the grand potential share the same eigenstates whereas their eigenvalues are related via a trivial shift
\begin{equation}
\Omega^{\text{A}}_{\mu} = E^{\text{A}}_{\mu} - \lambda \text{A} \, .
\end{equation}

The grand potential is thus partitioned according to 
\begin{equation}
\Omega\equiv \Omega_0+\Omega_1 \, ,
\end{equation}
such that the eigenstates of \(\Omega_0\) are known
\begin{subequations}
    \begin{align}
      \Omega_0 | \Phi \rangle   &\equiv  \Omega^{00} | \Phi \rangle  \, ,\\
     \Omega_0 | \Phi^{k_1k_2\cdots} \rangle   &\equiv  \left(\Omega^{00} + E_{k_1} + E_{k_2} + \cdots \right)| \Phi^{k_1k_2\cdots} \rangle\, .
    \end{align}
    \label{eq:eigenstatesH0}
\end{subequations}
The set of indices $\{k_1,k_2\cdots\}$ characterize the quasi-particle basis at play in the problem.  Correspondingly, excited eigen-energies of $\Omega_0$ involve the set of quasi-particle energies $\{E_{k_1},E_{k_2}\cdots\}$.

The degree of symmetry of $\Omega_0$, which can be lower than $\Omega$, dictates the characteristics of the quasi-particle indices. In Refs.~\cite{Tichai:2018eem,Tichai:2019ksh,Zare:2022cdw}, $\Omega_0$ and $|\Phi\rangle$ were enforced to be rotationally invariant, which is appropriate to doubly closed-shell and singly open-shell nuclei. In such a case, the indices characterizing the quasi-particle basis belong to the class defined through Eq.~\eqref{set1}. In the present work, $\Omega_0$ is allowed to break rotational symmetry while maintaining or not maintaining axial symmetry. In the former (latter) case, the indices characterizing the quasi-particle basis belong to the class defined through Eq.~\eqref{set2} (Eq.~\eqref{set3}).

\subsection{Wave operator}

Expansion methods connect the unperturbed state $|\Phi\rangle$ to the exact eigenstate $| \Psi^{\text{A}}_0 \rangle$ via the so-called {\it wave operator} \(\mathcal W\) 
\begin{equation}    
|\Psi^{\text{A}}_0 \rangle = \mathcal W |\Phi\rangle \, 
\label{eq:ManyBodyTensors}
\end{equation}
incorporating the effect of the residual interaction \(\Omega_1\). The goal is to expand $\mathcal W$ as a series to be truncated in order to approximate the exact ground state in a systematic way. The particular form employed to expand the wave operator characterizes the  method at play and the associated many-body tensors to be determined.

The first category concerns {\it perturbative} methods, \textit{i.e.} (B)MBPT~\cite{Ti20}, where the wave operator is parameterized according to
\begin{align}    
|\Psi^{\text{A}}_0 \rangle &= \sum_{p=0}^{\infty} \left(\left(\Omega^{00}-\Omega_0\right)^{-1} \Omega_1\right)^{p}_{\text{L}}|\Phi\rangle \nonumber \\
&\equiv \sum_{p=0}^{\infty} \sum_{q} \frac{1}{(2q)!} \sum_{k_1\cdots k_{2q}} C^{2q 0}_{k_1\cdots k_{2q}}(p)| \Phi^{k_1\cdots k_{2q}} \rangle \, ,
\label{eq:stateBMBPT}
\end{align}
where the subscript L denotes that only so-called {\it linked} terms must be retained~\cite{bloch58a}. In this context, the mode-$2q$ tensor $C^{2q 0}_{k_1\cdots k_{2q}}(p)$ expressed in the quasi-particle basis determines the coefficients of the $2q$ quasi-particle excitations emerging for any value of $p$ in the perturbative expansion. Working at a given perturbative order corresponds to truncating the sum over $p$ at a certain value\footnote{For each value of $p$ in Eq.~\eqref{eq:stateBMBPT}, the sum over $q$ naturally truncates at a finite value~\cite{Shavitt2009}. Eventually, the largest tensor mode $q_{\text{max}}$ is a function of the truncation order $p_{\text{max}}$.} $p_{\text{max}}$.

The second category concerns {\it non-perturbative} methods, \textit{e.g.} (Bogoliubov) coupled cluster ((B)CC)~\cite{SiDu15} theory relying on the ansatz
\begin{equation}    
|\Psi^{\text{A}}_0 \rangle = \text{e}^{\mathcal T_1 +\mathcal T_2 +\mathcal T_3 +\cdots} |\Phi\rangle \, ,
\label{eq:stateBCC}
\end{equation}
where the exponentiated connected $q$-tuple cluster operator $\mathcal T_{q}$ is represented in the quasi-particle basis by the mode-$2q$ tensor $t^{2q 0}_{k_1\cdots k_{2q}}$ that needs to be determined. In practice, a truncation is made that limits the set of cluster operators to $\{\mathcal T_q, 0\leq q\leq q_{\text{max}}\}$.

\subsection{Unknown tensors}

In both types of methods, the many-body tensors are functionals of the Hamiltonian\footnote{We refer equivalently to Hamiltonian or grand potential tensors given that they entertain a trivial connection.} tensors $\{\Omega^{ij}_{k_1\cdots k_{2k}}, i+j=2k, k=0,\cdots,k_{\text{max}}\}$ expressed in the {\it quasi-particle} basis via a basis transformation; see Sec.~\ref{hamtensorqpbasis} below. 

The simplicity of perturbative methods is that the associated tensors $C^{2q 0}_{k_1\cdots k_{2q}}(p)$ are {\it known} closed-form expressions in terms of the Hamiltonian tensors and quasi-particle energies $\{E_{k_l}\}$, \textit{i.e.} they can be computed explicitly at each order $p$ via a specific set of tensor networks of the form
\begin{equation}    
C^{2q 0}_{k_1\cdots k_{2q}}(p) = f^{\text{(B)MBPT}(p)}_{q}\left(\{\Omega^{ij}_{k_1\cdots k_{2k}}\}; \{E_{k_l}\}\right) \, .
\label{eq:TNBMBPT}
\end{equation}
Depending on the situation, the many-body tensors may have to be stored such that the memory footprint is either driven by the Hamiltonian tensor or by the many-body tensor of highest mode. The CPU cost is driven by the computation of Eq.~\eqref{eq:TNBMBPT} for the $q_{\text{max}}$ value corresponding to the truncation order $p_{\text{max}}$. Eventually, the reduction of the memory and CPU footprints requires the TF of the Hamiltonian tensors expressed in the quasi-particle basis given that such a TF propagates to the many-body tensors via Eq.~\eqref{eq:TNBMBPT}.

Non-perturbative methods constitute a greater challenge given that  many-body tensors such as $t^{2q 0}_{k_1\cdots k_{2q}}$ are only {\it implicit} functionals of the Hamiltonian tensors, \textit{i.e.} they are the solutions of a set of implicit tensor network equations for  $q \in [1,\ q_{\text{max}}]$
\begin{equation}    
0 = g^{(\text{BCCSD}\cdots)}_{q}\left(\{\Omega^{ij}_{k_1\cdots k_{2k}}\};\{t^{2q' 0}_{k_1\cdots k_{2q'}}\}\right) \, , 
\label{eq:TNBCC}
\end{equation}
with $q' \in [1,\ q_{\text{max}}]$. The coupled Eqs.~\eqref{eq:TNBCC} must be solved numerically in an iterative fashion, which typically requires to save several instances of the many-body tensors. Whereas the tensor $t^{2q_{\text{max}} 0}_{k_1\cdots k_{2q_{\text{max}}}}$ drives the storage footprint, Eqs.~\eqref{eq:TNBCC} for $q=q_{\text{max}}$ drives the CPU cost. Eventually, the TF must here be applied to the Hamiltonian tensors {\it and} employed as an ansatz for the unknown tensors such that the coupled Eq.~\eqref{eq:TNBCC} are transformed into implicit equations for the {\it factors} entering the TF~\cite{Schu17}.

\subsection{Observables}

Once the set of many-body tensors have been determined for the working truncation order, the ground-state expectation value of a given observable associated with the self-adjoint operator $F$ is computed via a specific tensor network combining the many-body tensors and the tensors defining $F$ in the quasi-particle basis. Taking the energy as an example and focusing on (B)MBPT, one has
\begin{equation}    
E^{\text{A}}_{0} =  \sum_{p=0}^{\infty} e^{(p)}_{\text{(B)MBPT}}\left(\{H^{ij}_{k_1\cdots k_{2k}}\};\{C^{2q 0}_{k_1\cdots k_{2q}}\}; \{E_{k_l}\}\right) \, , 
\label{eq:TNBMBPTobs}
\end{equation}
where the sum is to be truncated at perturbative order $p_{\text{max}}$ to deliver the approximation $E^{\text{A}}_{0} \approx E^{(p_{\text{max}})}_0$.

\section{(B)MBPT formalism}
\label{formalBMBPT}

The present work focuses on (B)MBPT(2) calculations of doubly open-shell nuclei in large bases. Working at second order in perturbation theory implies that $q_{\text{max}}=2$, \textit{i.e.} the largest many-body tensor in use is the mode-4 tensor $C^{40}_{k_1k_2k_3k_4}(2)$.

In the present section, the set of generic equations introduced in Sec.~\ref{sectMBtensors} are specified to both BMBPT(2) and MBPT(2). In the first case,  U(1) global gauge symmetry is broken whereas in the second case it is in fact enforced. Choosing one setting or the other has a significant impact on the characteristics of the quasi-particle indices and thus on the dimension of the tensors at play as will be discussed below. 

Furthermore, two different settings are considered regarding the breaking of rotational symmetry: either axial symmetry is enforced or it is allowed to break as well, leading to triaxially deformed calculations. The choice between these two options also impacts crucially the dimensions at play in the problem. 

Eventually, combining these different options define the four settings listed in Tab.~\ref{tab:methods}.

\subsection{BMBPT(2)}

\subsubsection{Bogoliubov reference state}

Obtaining the set of Bogoliubov quasi-particle creation and annihilation operators via a linear unitary transformation of the particle operators
\begin{equation}
    \begin{pmatrix}
        \beta\\
        \beta^\dag
    \end{pmatrix}
    \equiv
    \begin{pmatrix}
        U & V^* \\
        V & U^*
    \end{pmatrix}^\dag
    \begin{pmatrix}
        c \\c^\dag
    \end{pmatrix} \, ,
\end{equation}
the Bogoliubov reference state $|\Phi\rangle$ is introduced as the vacuum for this set, \textit{i.e.} $\beta_{k} \, | \Phi \rangle = 0$, $\forall \, k$.

\subsubsection{Hamiltonian tensors in qp basis}
\label{hamtensorqpbasis}

All operators can be re-expressed in normal order with respect to $| \Phi \rangle$. Starting from the rank-reduced version of $H$ defined in Eq.~\eqref{finalapproxH} and applying the normal-ordering procedure to the grand potential one obtains\footnote{Because $H$ and $\Omega$ only differ by the one-body particle number operator $A$, only their mode-0 and mode-2 components differ.}
\begin{equation}
\begin{aligned}
    \Omega &= \Omega^{00} \nonumber \\
    &+\Omega^{11} + \Omega^{20} + \Omega^{02} \nonumber \\
    &+ H^{22} + H^{31} + H^{13} + H^{40} + H^{04} \, , \label{OmegaQP}
\end{aligned}
\end{equation}
where, \textit{e.g.}, \(H^{ij}\) contains \(i\) qp creators and \(j\) qp annihilators and is represented by the mode-\((i+j)\) tensor $H^{ij}_{k_1\cdots k_{i+j}}$ expressed in the quasi-particle basis. For example, the component $H^{40}$ is represented by the mode-4 tensor $H^{40}_{k_1k_2k_3k_4}$ according to
\begin{equation}
    H^{40} \equiv \frac{1}{4!} \sum_{k_1k_2k_3k_4} H^{40}_{k_1k_2k_3k_4} \beta^\dag_{k_1}\beta^\dag_{k_2}\beta^\dag_{k_3} \beta^\dag_{k_4} \, . \label{H31qp}
\end{equation}

The tensors $H^{ij}_{k_1\cdots k_i k_{i+1} \cdots k_{i+j}} $ are fully anti-symmetric under the exchange of any pair among the $i$ ($j$) first (last) indices, \textit{i.e.}
\begin{equation}
H^{ij}_{k_1\cdots k_i k_{i+1} \cdots k_{i+j}} = \epsilon(\sigma_c) \epsilon(\sigma_a)  \, H^{ij}_{\sigma_c(k_1\cdots k_i) \sigma_a(k_{i+1} \cdots k_{i+j})} \, ,
\end{equation}
where $\epsilon(\sigma_c)$  ($\epsilon(\sigma_a)$) refers to the signature of the permutation $\sigma_c(\ldots)$ ($\sigma_a(\ldots)$) of the $i$ ($j$) indices corresponding to quasi-particle creation (annihilation) operators. 

These tensors are functionals of the Bogoliubov matrices $(U,V)$ and of the tensors defining the operators $H$ and $A$ in the initial sHO basis. For more details about the normal ordering procedure and for explicit expressions of $H^{ij}_{k_1\cdots k_i k_{i+1} \cdots k_{i+j}}$ up to $i+j=6$, see Refs.~\cite{Duguet:2015yle,Tichai18BMBPT,Arthuis:2018yoo,Ripoche2020}.

\subsubsection{Partitioning}

While it is not a necessity, present calculations employ a {\it canonical} Bogoliubov state $| \Phi \rangle$, \textit{i.e.} the unperturbed state is the Bogoliubov state solution of Hartree-Fock-Bogoliubov mean-field equations. This particular choice simplifies the tensor networks at play given that \(\Omega^{20}\) and \(\Omega^{02}\) vanish according to Brillouin's theorem. Furthermore, $\Omega^{11}$ is diagonal and delivers the (positive) quasi-particle energies such that
\begin{subequations}
\label{splittingBMBPT}
\begin{align}
    \Omega_0 &\equiv \Omega^{00} + \Omega^{11} \nonumber \\
    &=\frac{\langle \Phi | \Omega | \Phi \rangle}{\langle \Phi |  \Phi \rangle}  + \sum_{k} E_{k} \beta^{\dagger}_{k} \beta_{k} \, ,  \\
 \Omega_1 &\equiv H^{22} + H^{31} + H^{13} + H^{40} + H^{04} \, .
\end{align}
\end{subequations}
The excited eigenstates of $\Omega_0$ are obtained by applying an even number of quasi-particle excitation operators on the Bogoliubov vacuum
\begin{align}
| \Phi^{k_1k_2\cdots} \rangle   &\equiv  \beta^{\dagger}_{k_1} \beta^{\dagger}_{k_2}\cdots| \Phi \rangle\, .
\end{align}

\subsubsection{Many-body tensor in qp basis}

The second-order correction based on the canonical HFB reference state involves the sole mode-4 tensor $C^{40}_{k_1k_2k_3k_4}(2)$. The tensor network delivering it in terms of the Hamiltonian tensors (Eq.~\eqref{eq:TNBMBPT}) is given by~\cite{Duguet:2015yle,Arthuis:2018yoo}
\begin{subequations}
\label{tensorBMBPT2}
\begin{align}
C^{40}_{k_1k_2k_3k_4}(2) &= - \frac{H^{40}_{k_1k_2k_3k_4}}{E_{k_1}\!+\!E_{k_2}\!+\!E_{k_3}\!+\!E_{k_4}} \\
&= - H^{40}_{k_1k_2k_3k_4} D^{40}_{k_1k_2k_3k_4}\, , 
\end{align}
\end{subequations}
where the mode-$2q$ tensor $D^{2q0}_{k_1\cdots k_{2q}}\equiv (E_{k_1}\!+\! \cdots\!+\!E_{k_{2q}})^{-1}$ has been introduced.

\subsubsection{Energy correction}

The ground-state energy (Eq.~\eqref{eq:TNBMBPTobs}) is approximated in BMBPT(2) as 
\begin{align}
E^{\text{A}}_{0} &\approx E^{(2)}_0 \nonumber \\ 
&\equiv H^{00} + e^{(2)}_{\text{BMBPT}} \, ,
\end{align}
where the second-order correction is provided by the tensor network
\begin{subequations}
\label{eq:bmbpt_corr}
\begin{align} 
e^{(2)}_{\text{BMBPT}} &=  \frac{1}{4!}\sum_{k_1k_2k_3k_4} H^{04}_{k_1k_2k_3k_4}C^{40}_{k_1k_2k_3k_4}(2) \\
&= - \frac{1}{4!}\sum_{k_1k_2k_3k_4}
    \frac{|H^{40}_{k_1k_2k_3k_4}|^2}{E_{k_1}\!+\!E_{k_2}\!+\!E_{k_3}\!+\!E_{k_4}} \, \\
&= - \frac{1}{4!}\sum_{k_1k_2k_3k_4}  |H^{40}_{k_1k_2k_3k_4}|^2 D^{40}_{k_1k_2k_3k_4} \, .
\end{align}
\end{subequations}

\subsubsection{Dimensions}
\label{discussion_dim}

As visible from Eqs.~\eqref{tensorBMBPT2}-\eqref{eq:bmbpt_corr}, BMBPT(2) calculations require to deal with the two mode-4 tensors $H^{40}_{k_1k_2k_3k_4}$ and $D^{40}_{k_1k_2k_3k_4}$. At higher orders, additional Hamiltonian tensors are involved and the tensor networks at play become more complex.

\begin{figure*}
    \centering
    \includegraphics[width=1.0\textwidth]{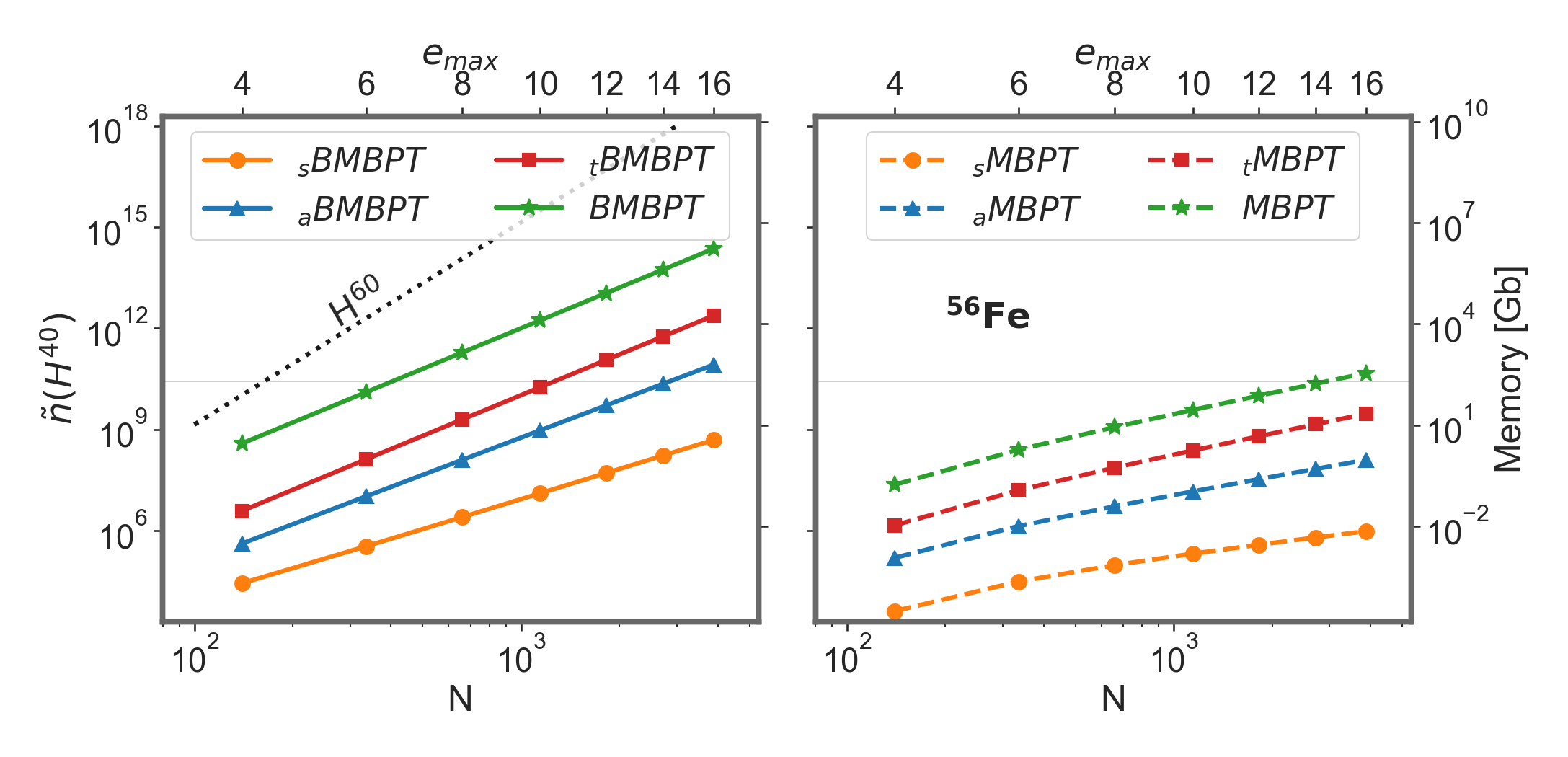}
    \caption{Size of the mode-4 tensor $H^{40}$ as a function of the one-body sHO basis size \(N\) in log-log scale. Left: tensor appropriate to BMBPT calculations. Right: tensor appropriate to MBPT calculations. The curves correspond to different storage schemes depending on the degree of symmetry at play in the calculation; see text for details. Gray horizontal lines highlight 200Gb of storage that constitutes the typical amount of memory available on one computational node. The dotted curve shows the naive storage requirement for the mode-6 tensor $H^{60}$.}
    \label{fig:scaling}
\end{figure*}

Using a fully naive approach, storing $H^{40}_{k_1k_2k_3k_4}$ scales as $N^4$ as is visible  for $^{56}$Fe from the curve (green stars) labeled as BMBPT in the left panel of Fig.~\ref{fig:scaling}. 

A first compression is obtained by exploiting the sparsity associated with parity and isospin conservations, \textit{e.g.} parity conservation imposes that $\pi_{k_1}\pi_{k_2}\pi_{k_3}\pi_{k_4}=1$, which means that half of the tensor entries vanish and thus do not have to be stored. The same factor is obtained from isospin conservation. Another compression by a factor $24$ is obtained from the antisymmetry (symmetry) of $H^{40}_{k_1k_2k_3k_4}$ ($D^{40}_{k_1k_2k_3k_4}$) under the exchange of any pair of indices. The overall compression by a factor 96 is effectively utilized for all cases under consideration\footnote{While not done in the present paper, the PAN@CEA numerical code  suite presently employed further authorizes to break parity symmetry. In such a case the compression factor is reduced to 48.}. 

Imposing that the unperturbed state $| \Phi \rangle$ is rotationally invariant as in Refs.~\cite{Tichai:2018eem,Tichai:2019ksh,Zare:2022cdw} allows one to employ J-coupled many-body tensors obtained via AMC techniques. Combined with (anti)symmetry and parity/isospin conservation, doing so reduces the storage of mode-4 tensors by about 6 orders of magnitude compared to the naive scheme as can be appreciated from the lowest curve labeled as $_s$BMBPT (orange circles) in the left panel of Fig.~\ref{fig:scaling}.

Allowing the Bogoliubov state $| \Phi \rangle$ to break rotational symmetry as is presently done, such a severe advantage cannot be exploited. Maintaining axial symmetry, as in $_\text{a}$BMBPT, $H^{40}_{k_1k_2k_3k_4}$ remains block diagonal with respect to the 2-body total angular momentum {\it projection} $M=m_{k_1}+m_{k_2}=-m_{k_3}-m_{k_4}$. Thus, while each index $k_i$ runs over the complete dimension $N$, an explicitly sparse character of the tensor can be employed\footnote{Effectively, the size reduction associated to the block diagonal structure with respect to $M$ is equivalent to saying that one out of the $4$ indices of the tensor run over $_\text{s}\tilde{N}$ rather than $N$.}. The compression at play corresponds to the curve labeled as $_\text{a}$BMBPT (blue triangles) in the left panel of Fig.~\ref{fig:scaling}.  Further breaking axial symmetry, as in $_\text{t}$BMBPT, no gain related to rotational symmetry can be taken advantage of such that the many-body tensors are expected to be dense. Only (anti)symmetry and parity/isospin conservation are left to be exploited, which corresponds to the curve labeled as $_\text{t}$BMBPT in the left panel of Fig.~\ref{fig:scaling}.

Eventually, the storage footprint of $_\text{s}$BMBPT(2) allows one to access semi-magic nuclei in large bases, \textit{i.e.} $e_\text{max}=16$, necessary to deal with heavy systems ($A\in [150,210]$)\footnote{As already discussed in Sec.~\ref{rankred}, such a mass regime is challenging for mode-6 tensors given that it typically requires $e_\text{3max}\in[24,28]$, see Refs.~\cite{Miyagi2021,Tichai:2023epe}.}. Dealing with doubly open-shell nuclei is much more challenging as can be seen from the left panel of Fig.~\ref{fig:scaling}. While BMBPT(2) calculations in axial symmetry can be handled up to rather large bases, \textit{i.e.} $e_\text{max}=12$, it is already problematic to compute triaxially deformed nuclei at $e_\text{max}=10$. This conclusion is all the more true for non-perturbative methods or when handling mode-6 tensors as the naive storage cost of $H^{60}$ shown in the left panel of Fig.~\ref{fig:scaling} demonstrates.

\subsection{MBPT(2)}

Whenever pairing correlations vanish or whenever U(1) global gauge symmetry is enforced, the Bogoliubov  state \(|\Phi\rangle\) reduces to a Slater determinant\footnote{Note that the Bogoliubov state solution of the HFB equations does not necessarily reduce to a Slater determinant in the zero-pairing limit. It only does so whenever the {\it naive filling} of the (canonical) single-particle shells corresponds to a closed-shell system. See Ref.~\cite{Duguet:2020hdm} for details.}. More specifically, the canonical reference state \(|\Phi\rangle\) is now the solution of Hartree-Fock (HF) mean-field equations.

The main feature of the present case relates to the fact that the quasi-particle basis naturally splits into two complementary sub-sets, \textit{i.e.} (i) the $A$ so-called {\it hole} indices (labeled by \(i,j,k\cdots\)) and the (ii) $N-A$ so-called {\it particle} indices (labeled by \(a,b,c\cdots\)). Eventually, the two components of the grand potential explicitly needed for the MBPT(2) application read now as 
\begin{subequations}
\label{newpartitioning}
\begin{align}
\Omega^{11} &\equiv \sum_{a} E_{a} \beta^{\dagger}_{a} \beta_{a} + \sum_{i} E_{i} \beta^{\dagger}_{i} \beta_{i} \, , \label{newpartitioninga} \\
    H^\text{40} &\equiv \frac1{2!2!} \sum_{abij} H^\text{40}_{abij}\beta^\dag_{i}\beta^\dag_{j}\beta^\dag_{b}\beta^\dag_{a} \, . \label{newpartitioningb}
\end{align}
\end{subequations}
Because of the explicit partitioning of the qp basis between particle and hole states, the antisymmetry of $H^\text{40}_{abij}$ only concerns indices belonging to the same partition. In Eq.~\eqref{newpartitioninga}, quasi-particle energies are further related to HF single-particle energies via 
\begin{subequations}
\label{qpsp}
\begin{align}
E_{a} &= |\epsilon_a - \lambda|=\epsilon_a - \lambda \, , \\
E_{i} &= |\epsilon_i - \lambda|=\lambda - \epsilon_i \, .
\end{align}
\end{subequations}

Eventually, the second-order many-body tensor reads now as
\begin{align}
C^{40}_{abij}(2) &= - \frac{H^\text{40}_{abij}}{\epsilon_a + \epsilon_b - \epsilon_i - \epsilon_j} \nonumber \\
&\equiv - H^\text{40}_{abij} D^\text{40}_{abij} \, ,  \label{tensorMBPT2}
\end{align}
such that the approximate ground-state energy is $E^{\text{A}}_{0} \approx  E^{(2)}_0 \equiv  H^{00} + e^{(2)}_{\text{MBPT}}$ with
\begin{subequations}
\label{eq:mbpt_corr}
\begin{align} 
e^{(2)}_{\text{MBPT}} &=  \frac{1}{2!2!}\sum_{abij} H^{04}_{abij}C^{40}_{abij}(2) \\
&= - \frac{1}{2!2!}\sum_{abij}
    \frac{|H^{40}_{abij}|^2}{\epsilon_a + \epsilon_b - \epsilon_i - \epsilon_j} \, \\
&= - \frac{1}{2!2!}\sum_{abij}  |H^{40}_{abij}|^2 D^\text{40}_{abij} \, .
\end{align}
\end{subequations}

Thanks to the partition of the qp basis, the size of the original tensors are significantly reduced in MBPT compared to BMBPT as can be seen by comparing the left and right panels of Fig.~\ref{fig:scaling}. The naive storage scheme of $H^{40}_{abij}$ scales as $A^2N^2$ instead of $N^4$ for $H^{40}_{k_1k_2k_3k_4}$ as can be inferred from the slope of the curves. While for small bases, \textit{e.g.} $e_\text{max}=4$, going from BMBPT back to MBPT corresponds to an initial compression by only about 1 order of magnitude in the storage footprint, for large basis sizes, \textit{e.g.} $e_\text{max}=16$, the initial compression reaches about 4 orders of magnitude in the naive approach and 3 orders of magnitude when exploiting (at least) (anti)symmetry. The latter step relates to the fact that (anti)symmetry only reduces the size of mode-4 tensors by a factor of 4 in MBPT rather than by a factor of 24 in BMBPT.

\section{Tensor factorization}
\label{SecTF}

Given the challenges set by the dimensions at play in (B)MBPT(2) calculations of mid-mass doubly open-shell nuclei discussed above in connection with Fig.~\ref{fig:scaling}, the goal is to provide a factorized representation of the mode-4 Hamiltonian and many-body tensors at play, \textit{e.g.} $H^{40}_{k_1k_2k_3k_4}$ and $C^{40}_{k_1k_2k_3k_4}(2)$\footnote{The factorization of $C^{40}_{k_1k_2k_3k_4}(2)$ is achieved by factorizing separately $H^{40}_{k_1k_2k_3k_4}$ and $D^{40}_{k_1k_2k_3k_4}$. This option is not mandatory and the employed TF technique could be applied to $C^{40}_{k_1k_2k_3k_4}(2)$ at once.}. The goal is to reduce significantly the memory and CPU footprints while inducing a negligible error on physical observables, \textit{e.g.} the ground-state energy.

Ideally, the employed TF must fulfill the following requirements
\begin{enumerate}
    \item It must be inexpensive to compute and memory efficient.
    \item The initial tensor should not be constructed.
    \item Its truncation must lead to a significant compression.
    \item The impact of its truncation must be monitored
\begin{enumerate}
    \item in an adaptive fashion,
    \item without comparing to the full tensor.
\end{enumerate}
\end{enumerate}

\subsection{Error and compression factor}

The relative error of the approximation $\tilde{T}$ to the initial tensor $T$ is quantified here as
\begin{equation}\label{eq:tf_err}
    \epsilon \equiv \frac{\|T-\tilde T\|_\text{F}}{\|T\|_\text{F}} \, ,
\end{equation}
where  the Frobenius norm of a mode-\(k\) tensor \(T\in\mathbb R^{n_1\times\cdots\times n_k}\) is given by
\begin{equation}
    \|T\|_\text{F}\equiv \sum_{i_1\cdots i_k} T_{i_1\cdots i_k}^2 \, .
\end{equation}

Defining as \(n(T)\) the number of elements needed to specify an arbitrary tensor $T$, the compression factor achieved via an approximation \(\tilde T\) is given by
\begin{equation}\label{eq:tf_tc}
    t_\text{c} \equiv \frac{n(\tilde T)}{n(T)} \, .
\end{equation}

\subsection{BMBPT(2)}

\subsubsection{$D^{40}_{k_1k_2k_3k_4}$}
\label{secEdenominator}

The mode-4 tensor $D^{40}_{k_1k_2k_3k_4}$ capturing the energy denominator in  Eqs.~\eqref{tensorBMBPT2} and~\eqref{eq:bmbpt_corr} is factorized using the Laplace transform~\cite{Braess05,tichai2019tf}
\begin{equation}
    \tilde{D}^{40}_{k_1k_2k_3k_4} \equiv \sum_{i=1}^{n_\text{d}} d^i_{k_1} d^i_{k_2} d^i_{k_3} d^i_{k_4} \, ,
\end{equation}
which, for a small number \(n_\text{d}\approx10\) of vectors \(d^i\), is essentially exact.

\subsubsection{$H^{40}_{k_1k_2k_3k_4}$ in $_\text{t}$BMBPT}

In $_\text{t}$BMBPT, the mode-4 tensor $H^{40}_{k_1k_2k_3k_4}$ in the qp basis is dense and thus a good candidate for a singular value decomposition (SVD). The tensor is put in matrix form
\begin{equation}
{\correction{A}{W}}_{(k_1k_2)(k_3k_4)} \equiv  H^{40}_{k_1k_2k_3k_4}  \, ,
\end{equation}
by grouping indices two-by-two before a rank-\(r_\text{c}\) truncated SVD\footnote{Properties of $H^{40}_{k_1k_2k_3k_4}$ ensures that \(\correction{A}{W}\) is a symmetric matrix such that a symmetric eigen-decomposition is in fact performed instead of a SVD as detailed in~\ref{appsymdiago}. This allows one to store only one set of singular (\textit{i.e.} eigen) vectors.} is performed to obtain the approximation to $\correction{A}{W}$ as
\begin{equation}
\tilde {\correction{A}{W}}_{(k_1k_2)(k_3k_4)} \equiv \sum_{\mu=1}^{r_\text{c}} s_\mu C_{(k_1k_2)}^\mu D_{(k_3k_4)}^\mu \, , \label{centralSVD}
\end{equation}
\(r_\text{c}\) being referred to as the central rank. At this stage, the number of stored elements is
\begin{align}
n(\tilde {\correction{A}{W}}) &= n(s) + n(C) + n(D) \nonumber \\
        &= r_\text{c} (1 + N^2) \, .
\end{align}

The left (right) \(C\) (\(D\)) singular vectors can be seen as mode-3 tensors that are  expected to be dense and thus amenable to a second SVD on each mode-2 \(C^\mu\) (\(D^\mu\)) component\footnote{\label{foot:it}An alternative option would be to design an approximate sparse format by only keeping entries lower than a given threshold. This combination of TF and IT~\cite{tichai2019pre,Porro:2021rjp} techniques is left to a future work.}. Focusing on \(C^\mu\) as an example and exploiting that such a matrix is antisymmetric\footnote{Exploiting explicitly the antisymmetry of the matrices reduces the memory cost by a factor of two.}, one obtains
\begin{equation}\label{eq:C_decomp}
\sqrt{s_\mu}\tilde C_{(k_1k_2)}^\mu \equiv \sum_{i=0}^{r^{\mu}_\text{s}/2} \lambda^{\mu}_i (X_{k_1}^{\mu 2i} X_{k_2}^{\mu 2i+1}  - X_{k_1}^{\mu 2i+1} X_{k_2}^{\mu 2i} ) \, .
\end{equation}
Clearly, the rank $r^\mu_\text{s}$ depends on \(\mu\) whenever employing a fixed threshold on the singular values. For estimation purposes, it is useful to introduce the mean rank $r_\text{s}$. Given that the antisymmetric SVD only requires to store one set of singular vectors, the number of elements stored after this second stage is 
\begin{align}
n(\tilde {\correction{A}{W}}) &= r_\text{c} (1 + r_\text{s}+ r_\text{s} N) \, .
\end{align}

\subsubsection{Correlation energy}

The second-order correlation energy defined in Eq.~\eqref{eq:bmbpt_corr} can be re-expressed using the TF. Defining intermediates
\begin{subequations}
\label{eq:LR_intermediate}
\begin{align}
    L(i,\mu,\nu) &\equiv \sqrt{s_\mu s_\nu} \sum_{k_1k_2}
    d^i_{k_1}
    d^i_{k_2}
    \tilde C^\mu_{k_1k_2} \tilde C^{\nu*}_{k_1k_2} \ ,\\
    R(i,\mu,\nu) &\equiv \sqrt{s_\mu s_\nu}\sum_{k_1k_2}
    d^i_{k_1}
    d^i_{k_2}
    \tilde D^\mu_{k_1k_2} \tilde D^{\nu*}_{k_1k_2} \ ,
\end{align}
 \end{subequations}
the correlation energy can be rewritten as
\begin{equation}
    e^{(2)}_{\text{BMBPT}}\approx-\frac1{4!}\sum_{i=1}^{n_\text{d}}\sum_{\mu=1}^{r_\text{c}}\sum_{\nu=1}^{r_\text{c}}
    L(i,\mu,\nu) R(i,\mu,\nu) \, . \label{energycorrectTF}
\end{equation}

In cases where \(r_s^\mu\) and \(r^\nu_s\) are small, it is interesting to reshuffle sums and build new intermediates in the computation of $L$ by explicitly exploiting Eq.~\eqref{eq:C_decomp} instead of reconstructing \(\tilde C^\mu\) on the fly to gain computational time\footnote{The same discussion naturally applies to $\tilde D^\mu$ in order to re-express \(R\).}. Defining
\begin{equation}
    K^{i\mu \nu }_{ss'}\equiv \sum_{k_1} d^i_{k_1} X^{\mu s*}_{k_1}X^{\nu s'}_{k_1},
\end{equation}
\(L\) is advantageously re-expressed as
\begin{align}
    L(i,&\mu,\nu) = 2 \sum_{j=0}^{r^\mu_s/2}\sum_{j'=0}^{r^\nu_s/2} \lambda^\mu_j\lambda^\nu_{j'} \\&\times\left( 
     K^{i\mu\nu}_{2j2j'}K^{i\mu\nu}_{(2j+1)(2j'+1)} -
     K^{i\mu\nu}_{2j(2j'+1)}K^{i\mu\nu}_{(2j+1)2j'} \right) .\nonumber
\end{align}

\subsubsection{$H^{40}_{k_1k_2k_3k_4}$ in $_\text{a}$BMBPT}

In $_\text{a}$BMBPT, ${\correction{A}{W}}_{(k_1k_2)(k_3k_4)}$ is block diagonal with respect to $M=m_{k_1}+m_{k_2}=-m_{k_3}-m_{k_4}$. The associated sparsity discussed in Sec.~\ref{discussion_dim} is exploited from the outset such that the SVD in Eq.~\eqref{centralSVD} is applied to each sub-block separately. 

Contrary to $_\text{t}$BMBPT, matrices $C^\mu$ and $D^\mu$ resulting from the first SVD carry a symmetry quantum number $M_\mu=m_{k_1}+m_{k_2}$ and are thus sparse by construction. Subsequently, it is more efficient to exploit this sparsity explicitly rather than performing a second SVD on the sparse matrices. Effectively, this corresponds to storing the final number of elements
\begin{align}
n(\tilde {\correction{A}{W}}) &= r_\text{c} (1 + N{}_\text{s}\tilde N) \, .
\end{align}
Eventually, the second-order correction to the energy is computed according to Eqs.~\eqref{eq:LR_intermediate}-\eqref{energycorrectTF} without any approximation on the set of sparse $C^\mu$ and $D^\mu$ matrices.

\subsubsection{Randomized SVD and adaptative range-finder}

A key aspect of the present work relates to the possibility to perform the central SVD without ever constructing the matrix ${\correction{A}{W}}$ explicitly given that its dimension can be prohibitive to begin with as discussed in Sec.~\ref{discussion_dim}. 

This objective can be achieved thanks to a randomized implementation of the truncated SVD~\cite{martinsson2021randomized,tropp2023randomized,Tichai:2023dda} (RSVD) that only requires to evaluate matrix-vector products.

The first step consists of finding an approximate basis for the \emph{range} of \({\correction{A}{W}}\), {\it \textit{i.e.}} the subspace spanned by its column vectors. This step, referred to as the range-finder procedure, is followed by an actual SVD of the matrix projected onto this subspace before truncating it according to the rank \(r_\text{c}\). See \ref{RandSVD} for details on the RSVD procedure.

The main challenge resides in the choice of the range-finder algorithm that must be fast to compute and economical, such that \(r_\text{c}\) can eventually be taken as small as possible. The blocked Lanczos range-finder~\cite{martinsson2021randomized} presently utilized is detailed in~\ref{adaptrange-finder}.

Furthermore, the aim is to  have an adaptive procedure to determine the optimal rank \(r_\text{c}\) {\it a posteriori}, {\it given} a target precision \(\epsilon\). In order to achieve this goal, a stochastic estimator \(\epsilon_\text{c}\) of \(\epsilon\) is employed that is much cheaper to compute and does not require to build (and store) the original matrix ${\correction{A}{W}}$. This estimator is also described in~\ref{adaptrange-finder}.

\subsubsection{Implicit product}

The RSVD only requires the knowledge of the matrix-vector product \(X\rightarrow {\correction{A}{W}}X\) to perform a decomposition of \({\correction{A}{W}}\). It is particularly interesting to evaluate this product without explicitly computing \({\correction{A}{W}}\). The same motivation is at the origin of the finite amplitude method (FAM) approach to the quasi particle random phase approximation (QRPA). Indeed, in the FAM method the QRPA equations are solved without actually evaluating the QRPA matrix~\cite{Nakatsukasa07a,Carlsson_2012}.

The presently developed algorithm follows the same route to evaluate the needed product of the matrix ${\correction{A}{W}}$ and a random vector $X^{02}$
\begin{equation}
    X^{02}\rightarrow Y^{20}_{k_1k_2} = \sum_{k_3k_4} H^{40}_{k_1k_2k_3k_4} X^{02}_{k_3k_4},
\end{equation}
without actually working in the qp basis and thus without needing $H^{40}_{k_1k_2k_3k_4}$ to begin with. The corresponding algorithm, detailed in Alg.~\ref{alg:bogo-contraction}, proceeds via a back-and-forth transformation to the sHO basis where AMC can be exploited such that the dimension at play in the explicit matrix-vector product is easily handled. This procedure is a key feature of the present work.

\begin{algorithm}
\caption{Matrix-vector product in BMBPT(2)\label{alg:bogo-contraction}}
    \KwData{\( X^{02}_{k_3k_4}\) a mode-2 antisymmetric tensor in qp basis.}
    \KwResult{\( Y^{20}_{k_1k_2}\equiv\sum_{k_3k_4} H^{40}_{k_1k_2k_3k_4} X^{02}_{k_3k_4}\)}
    
    *\ \ \ Transform \(X^{02}\) to sHO basis
    
    \(\bar \rho \gets  V^*  X^{02}  U^\dagger\),
    \(\bar \kappa \gets  V^*  X^{02} V^\dagger\),
    \(\bar \kappa'^{\color{red}*}\gets   U^*  X^{02}  U^\dagger\).

    *\ \ \  Contract with J-coupled $H$ tensor in sHO basis.
    
    \(\bar h_{\alpha\gamma}\gets 2 \sum_{\beta\delta} H_{\alpha\beta\gamma\delta}\cdot \bar\rho_{\delta\beta}\), 
    
    \(\bar\Delta_{\alpha\beta}\gets \sum_{\gamma\delta} H_{\alpha\beta\gamma\delta}\cdot \bar\kappa_{\gamma\delta}\), 
    
    \(\bar \Delta'_{\alpha\beta}\gets\sum_{\gamma\delta} H_{\alpha\beta\gamma\delta}\cdot \bar\kappa'_{\gamma\delta}\).

    *\ \ \  Extract 20 component back in qp basis

    \( Y^{20}\gets - U^\dagger \bar h V^*
    +  V^\dagger \bar h^\transpose  U^*
    -  U^\dagger \bar \Delta   U^*
    -  V^\dagger \bar \Delta'^{\color{red}*}  V^*\)

    \KwReturn {\( Y^{20}\)}
\end{algorithm}

\subsection{MBPT(2)}

Even though the TF works similarly in MBPT(2) compared to BMBPT(2), a few key differences must be highlighted.

While the tensor $D^{40}_{abij}$ entering Eqs.~\eqref{tensorMBPT2} and~\eqref{eq:mbpt_corr} is factorized to very good accuracy following the approach already described in Sec.~\ref{secEdenominator}, the central SVD of $H^{40}_{abij}$ presents more freedom as far as the way the tensor can be transformed into a matrix. Indeed, one can proceed according to either
\begin{subequations}
\begin{equation}
{\correction{A}{W}}_{(ij),(ab)} \equiv  H^{40}_{abij}  \, ,
\end{equation}
or
\begin{equation}
{\correction{A}{W}}_{(ia),(jb)} \equiv H^{40}_{abij}  \, .
\end{equation}
\end{subequations}
The first option leads to an ill-balanced matrix of dimensions \(A^2\times(N-A)^2\) while the second yields a square matrix of size \(A(N-A)\times A(N-A)\). Because the compression  provided by the truncated SVD is much more efficient for well-balanced matrices, the second option is chosen. The truncated SVD delivers
\begin{equation}
    \tilde {\correction{A}{W}}_{(ia)(jb)}  \equiv \sum_{\mu=1}^{r_\text{c}} s_\mu C_{(ia)}^\mu D_{(jb)}^\mu \, ,
\end{equation}
such that the numbers of elements to store is
\begin{equation}
    n(\tilde {\correction{A}{W}}) = r_\text{c} (1 + 2A(N-A))\, .
\end{equation}

The resulting matrices \(C^\mu\) and \(D^\mu\) are very unbalanced with dimensions \(A\times (N-A)\) such that performing a second SVD brings no further benefit. Of course, in $_\text{a}$MBPT the sparse format of \(C^\mu\) and \(D^\mu\) associated with their diagonality in $M_\mu$ can still be exploited to compress the data.

The implicit product necessary to perform the RSVD
\begin{equation}
    X\rightarrow Y_{ia} = \sum_{jb} H^{04}_{abij} X_{jb},
\end{equation}
is performed following the simpler algorithm detailed in Alg.~\ref{alg:sd-contraction}.

\begin{algorithm}
\caption{Matrix-vector product in  MBPT(2)\label{alg:sd-contraction}}
    \KwData{\( X_{jb}\) a mode-2 tensor in qp basis}
    \KwResult{\(Y_{ia} =\sum_{jb} H^{40}_{abij} X_{jb}\)}

    *\ \ \  Transform \(X\) to sHO basis
    
    \(\bar \rho \gets  V^*  X  {U}^\dagger\),

    *\ \ \   Contract with J-coupled $H$ tensor in sHO basis
    
    \(\bar h_{\alpha\gamma}\gets \sum_{\beta\delta} H_{\alpha\beta\gamma\delta}\cdot \bar\rho_{\delta\beta}\), 

    *\ \ \  Transform back to qp basis

    \( Y\gets -  {V}^\dagger \bar h^\transpose U^*
    \)

    \KwReturn {\( Y\)}
\end{algorithm}

Eventually, the approximation to the second-order energy correction $e^{(2)}_{\text{MBPT}}$ is computed according to Eqs.~\eqref{eq:LR_intermediate}-\eqref{energycorrectTF} except for the prefactor that is $(2!2!)^{-1}$ rather than $(4!)^{-1}$.

\subsection{Complexity analysis}

\begin{table*}
    \centering
\begin{tabular}{r|c|c|c}
    \hline
    Setting & Memory & Construction & \(e^{(2)}\) \\
    \hline
    \hline
    $_\text{t}$BMBPT(2)  & $O(N^4)$ & $O(N^5)$ & $O(N^4)$ \\
    \hline
    svd\correction{$^{2}$}{}-$_\text{t}$BMBPT(2)   &  $O(r_\text{c} r_\text{s} N)$ & $O(r_\text{c} N^4)$ & $O(n_\text{d} r_\text{c}^2 N^2)$ \\
    \hline
    \hline
    $_\text{a}$BMBPT(2) & $O(_\text{s}\tilde N N^3)$ & $O(_\text{s}\tilde N N^4)$ & $O(_\text{s}\tilde N N^3)$ \\ 
    \hline
    svd-$_\text{a}$BMBPT(2)  & $O(r_\text{c} {}_\text{s}\tilde N N)$ & $O(r_c{}_\text{s}\tilde N^2 N^2) $ & $O(n_dr_c^2 {}_\text{s}\tilde N N)$ \\
    \hline
    \hline
    $_\text{t}$MBPT(2) & $O(A^2N^2)$ & $O(A N^4)$ & $O(A^2N^2)$ \\
    \hline
    svd-$_\text{t}$MBPT(2)  & $O(r_\text{c} AN)$ & $O(r_\text{c} A^2N^2)$ & $O(n_\text{d} r_\text{c}^2 AN)$ \\
    \hline
    \hline
    $_\text{a}$MBPT(2) & $O(A^2\ _\text{s}\tilde N N)$ & $O(A \ _\text{s}\tilde N N^3)$ & $O(A^2\ _\text{s}\tilde N N)$ \\
    \hline
    svd-$_\text{a}$MBPT(2)  & $O(r_\text{c} A\ _\text{s}\tilde N)$ & $O(r_{\text c}A^2\ _\text{s}\tilde N N)$ &
    $O(n_\text{d} r_\text{c}^2 A\ _\text{s}\tilde N)$\\
    \hline
    \end{tabular}
    \captionof{table}{Complexity of original and TF (B)MBPT(2) calculations in the four settings of present interest. }
    \label{tab:complexity}
\end{table*}

Memory and operation complexities are easily expressed as a function of \(N, \, _\text{s}\tilde{N}, A, r_\text{c}, r_\text{s}, n_\text{d}\) and are reported for all cases of interest in Tab.~\ref{tab:complexity}. Enforcing axial symmetry typically corresponds to replacing one factor $N$ by a factor $_\text{s}\tilde{N}$.

Note that the memory requirements displayed in Tab.~\ref{tab:complexity} suppose that both in the exact and TF cases, the many-body tensor is indeed stored. While it is not strictly necessary to compute the correlation energy, it becomes necessary to access the one- and two-body density matrices from which any one- and two-body observables can be calculated.

\section{Results}
\label{sec:results}

\begin{figure*}
    \centering
    \includegraphics[width=\textwidth]{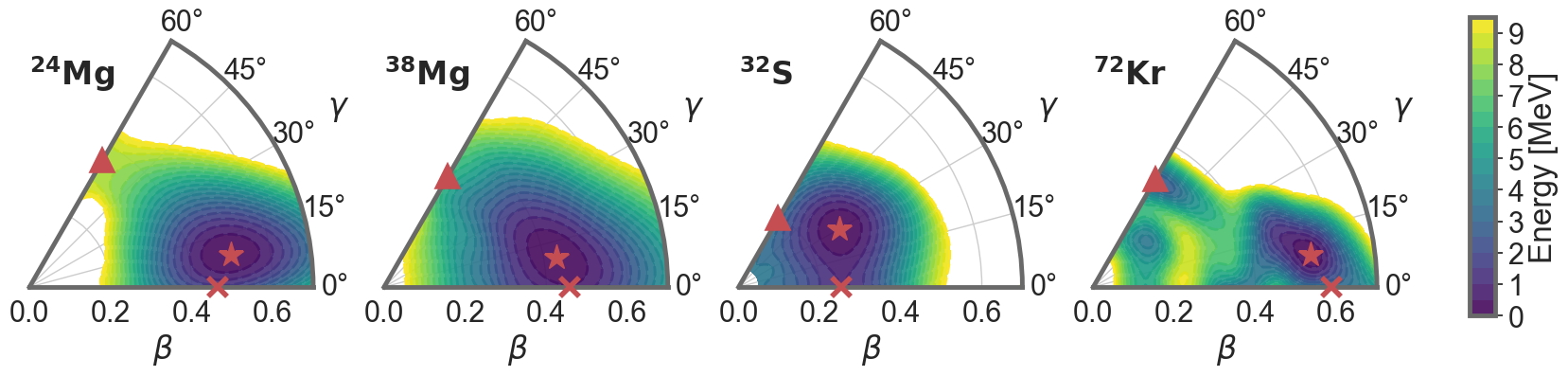}
    \caption{\label{fig:hfb-pes} HFB total energy surface of selected systems in the \((\beta,\gamma)\) deformation plane. Prolate, oblate and triaxial mimina of the energy surface are marked with red cross, triangle and star respectively. $e_{\!_{\;\text{max}}}=12$ is employed here.}
\end{figure*}

\subsection{Numerical setting}

The TF is presently tested for $_\text{a,t}$(B)MBPT(2) calculations of doubly open-shell $^{24}$Mg, $^{38}$Mg, $^{32}$S and $^{72}$Kr, as well as of $^{56}$Fe at the end of the paper.

The sHO one-body basis characterized by the frequency $\hbar\omega=12$~MeV is employed with $e_{\!_{\;\text{max}}}$ ranging from 4 to 16, \textit{i.e.} $N$ ranging from 140 to 3876. The representation of three-body operators is restricted by employing three-body states up to $e_{\!_{\;\text{3max}}}=16$.

Calculations are performed using the EM1.8/2.0 Hamiltonian~\cite{Hebeler11a} containing  two- (2N) and three-nucleon (3N) interactions. Matrix elements were generated using the Nuhamil code~\cite{Miyagi2023}. The three-body force is approximated via the rank-reduction method developed in Ref.~\cite{Frosini21a}. While the quantitative results may depend on the particular Hamiltonian under use, the TF presently employed is generic and can later on be tested in conjunction with any Hamiltonian. 

\subsection{Reference state}

The HFB total energy surface in the \((\beta,\gamma)\) deformation plane of each of the four nuclei under consideration is depicted in Fig.~\ref{fig:hfb-pes}. 

The four nuclei exhibit a triaxially deformed minimum (red star) along with axial prolate  (red cross) and oblate (red triangle) saddle points. The subsequent perturbative calculation performed on top of the absolute minimum calls for the $_\text{t}$(B)MBPT setting whereas using the saddle points as reference states allows one to use the simpler $_\text{a}$(B)MBPT one. 

The reference states located at triaxial absolute minima happen to be unpaired  with the EM 1.8/2.0 Hamiltonian, \textit{i.e.} they reduce to HF states such that one can limit oneself to using $_\text{t}$MBPT rather than $_\text{t}$BMBPT. Targeting other triaxially deformed nuclei, using different Hamiltonians and/or performing calculations {\it constrained} on pairing may lead to paired triaxial minima such that $_\text{t}$BMBPT remains potentially pertinent. In the present study, the $_\text{t}$BMBPT setting will be tested based on oblate reference states (red triangles) for which particle number symmetry is spontaneously broken.

\subsection{MBPT(2)}

\subsubsection{Prolate reference state}

\begin{figure}
    \centering
    \includegraphics[width=.52\textwidth]{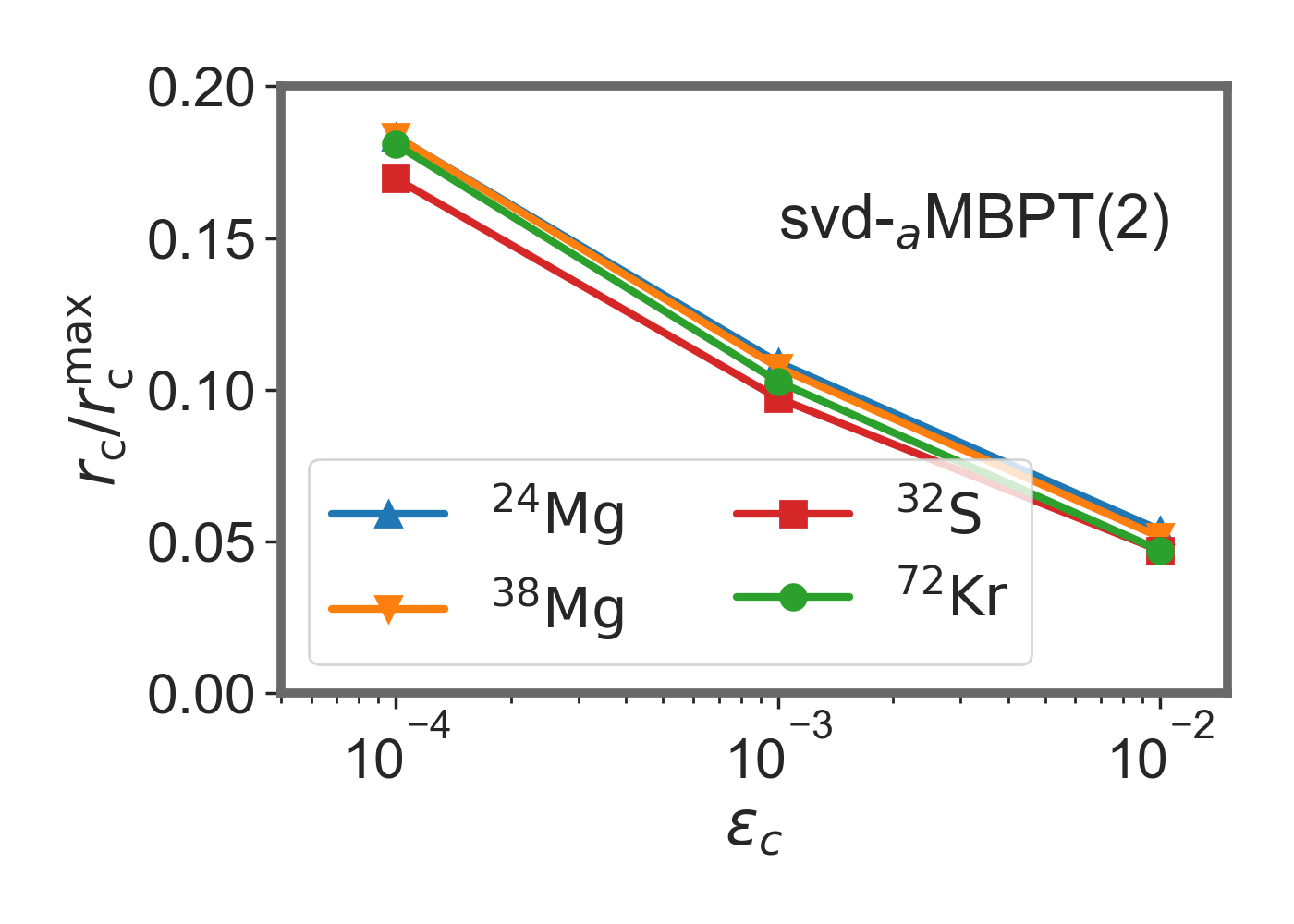}
    \caption{\label{fig:r-mbpt}Ratio between central rank \(r_\text{c}\) and theoretical upper bound \(r_\text{c}^\text{max}=A(N-A)=A(1820-A)\) as a function of the RSVD truncation error \(\epsilon_\text{c}\) in svd-$_\text{a}$MBPT(2) calculations. Calculations are performed for \(e_\text{max}=12\) (corresponding to \(N=1820\)).}
\end{figure}

The TF is first tested based on unpaired prolate reference states (red crosses in Fig.~\ref{fig:hfb-pes}) using a one-body sHO basis corresponding to $e_{\!_{\;\text{max}}}=12$ ($N=1820$). 

Employing $_\text{a}$MBPT(2), the ratio of the rank \(r_\text{c}\) of the central RSVD to its theoretical upper bound (keeping all singular values) \(r_\text{c}^\text{max}=A(N-A)\) is shown in Fig.~\ref{fig:r-mbpt} to decrease as a function of the (stochastically estimated) RSVD tolerance \(\epsilon_\text{c}\). The trend is seen to be essentially nucleus-independent. For a tolerated error of 0.01\%, only one sixth of the singular vectors of the initial tensor need to be kept, thus leading to a decent compression.

\begin{figure}
    \centering
    \includegraphics[width=.48\textwidth]{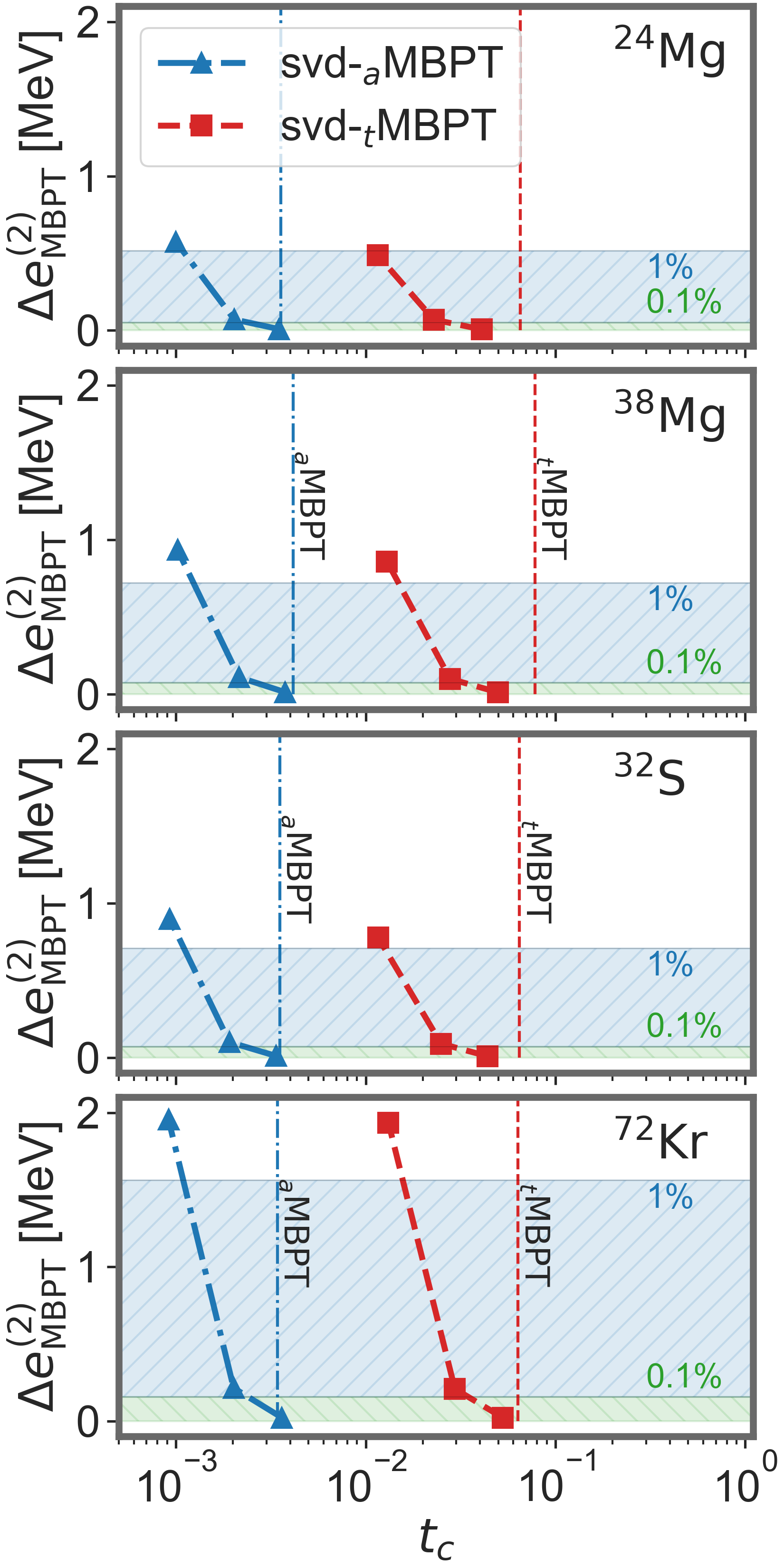}
    \caption{\label{fig:tfpt-bench} Error on $e^{(2)}_{\text{MBPT}}$ based on unpaired prolate reference states in svd-$_\text{a}$MBPT(2) (blue triangles) and svd-$_\text{t}$MBPT(2) (red squares) calculations against the compression factor \(t_\text{c}\) computed with respect to the naive storage scheme of $H^{40}_{abij}$. Points correspond to three values of the RSVD tolerance \(\epsilon_\text{c}=\{1\%, 0.1\%, 0.01\%\}\) from left to right and are linked by lines to guide the eye. The compression factor corresponding to the optimal storage scheme in $_\text{a}$MBPT(2) ($_\text{t}$MBPT(2)) is shown as a vertical blue dashed line (red dashed-dotted line). Regions lying within 1\% resp. (0.1\%) of the exact $e^{(2)}_{\text{MBPT}}$ are highlighted. Calculations are performed for \(e_\text{max}=12\). }
\end{figure}

This advantage is now characterized employing the compression factor $t_\text{c}$ computed with respect to the {\it naive} storage scheme of $H^{40}$, \textit{i.e.} $A^2N^2$. In Fig.~\ref{fig:tfpt-bench}, the error on $e^{(2)}_{\text{MBPT}}$ is thus plotted against $t_c$ for svd-$_\text{a}$MBPT(2) and svd-$_\text{t}$MBPT(2) using three values of the central RSVD tolerance \(\epsilon_\text{c}=\{1\%, 0.1\%, 0.01\%\}\). Vertical lines show the compression factors corresponding to the optimal storage schemes in $_\text{a}$MBPT(2) and $_\text{t}$MBPT(2) (blue triangles and red squares in the right panel of Fig.~\ref{fig:scaling}, respectively) calculations.

While the compression is significant against the naive storage scheme in both settings, the TF becomes favorable against the corresponding initial optimal storage scheme for tolerated errors larger than \(0.01\%\). The reason why it is unfavorable for errors smaller than \(0.01\%\) relates to the fact that the advantage due to the partial antisymmetry of $H^{40}_{abij}$ in the initial storage scheme is partially lost\footnote{The conservation of isospin allows to perform the TTF separately on each isospin components and partially exploits antisymmetry.} in the TF format in MBPT\footnote{It happens that in view of applying the TF format in higher-order (B)MBPT or in non-perturbative calculations, storing matrix elements without making explicit use of such a symmetry allows one to better align operations and take full advantage of modern CPU/GPU architectures as well as optimized linear algebra libraries. Consequently, such a feature may help accelerate some operations by avoiding permutations.}. Eventually, the TF is shown to display the following features:
\begin{itemize}
    \item The relative error on $e^{(2)}_{\text{MBPT}}$ closely follows the value of \(\epsilon_\text{c}\). It suggests that the error on observables is directly deducible from the error on the central RSVD that is monitored by the user through the chosen value of \(\epsilon_\text{c}\). This result is key to eventually characterize the results based on TTF without having the un-approximated calculation at hand. 
    \item The approximate energy delivered through TTF approaches the reference value from \correction{below}{above}, thus allowing lower bound estimates.
    \item In absolute, the memory cost of $_\text{a}$MBPT(2) is about one order of magnitude smaller than for $_\text{t}$MBPT(2) due to the sparse format associated with axial symmetry. The compression factor offered by the TTF is however similar in both cases, \textit{i.e.} of the order of $0.2$ ($0.4$) for a tolerated error of  \(1\%\) (\(0.1\%\)) at \(e_\text{max}=12\). As shown later on, the gain provided by TTF increases for even larger \(e_\text{max}\) values.
    \item The above conclusions are valid for the four nuclei under study  spanning the mass range $A\in [24,72]$.
\end{itemize}

\subsubsection{Triaxial calculations in large bases}

Eventually, one objective of the present study is to perform calculations based on triaxial reference states in large bases that cannot be done without TF. In the present section, svd-$_\text{t}$MBPT(2) is tested for both prolate and triaxial reference states while increasing the basis size to  \(e_\text{max}=14\) ($N=2720$). The tolerance is fixed to \(\epsilon_\text{c}=0.01\%\), which ensures a similar error on the second-order correlation energy.  

\begin{figure}
    \centering
    \includegraphics[width=.5\textwidth]{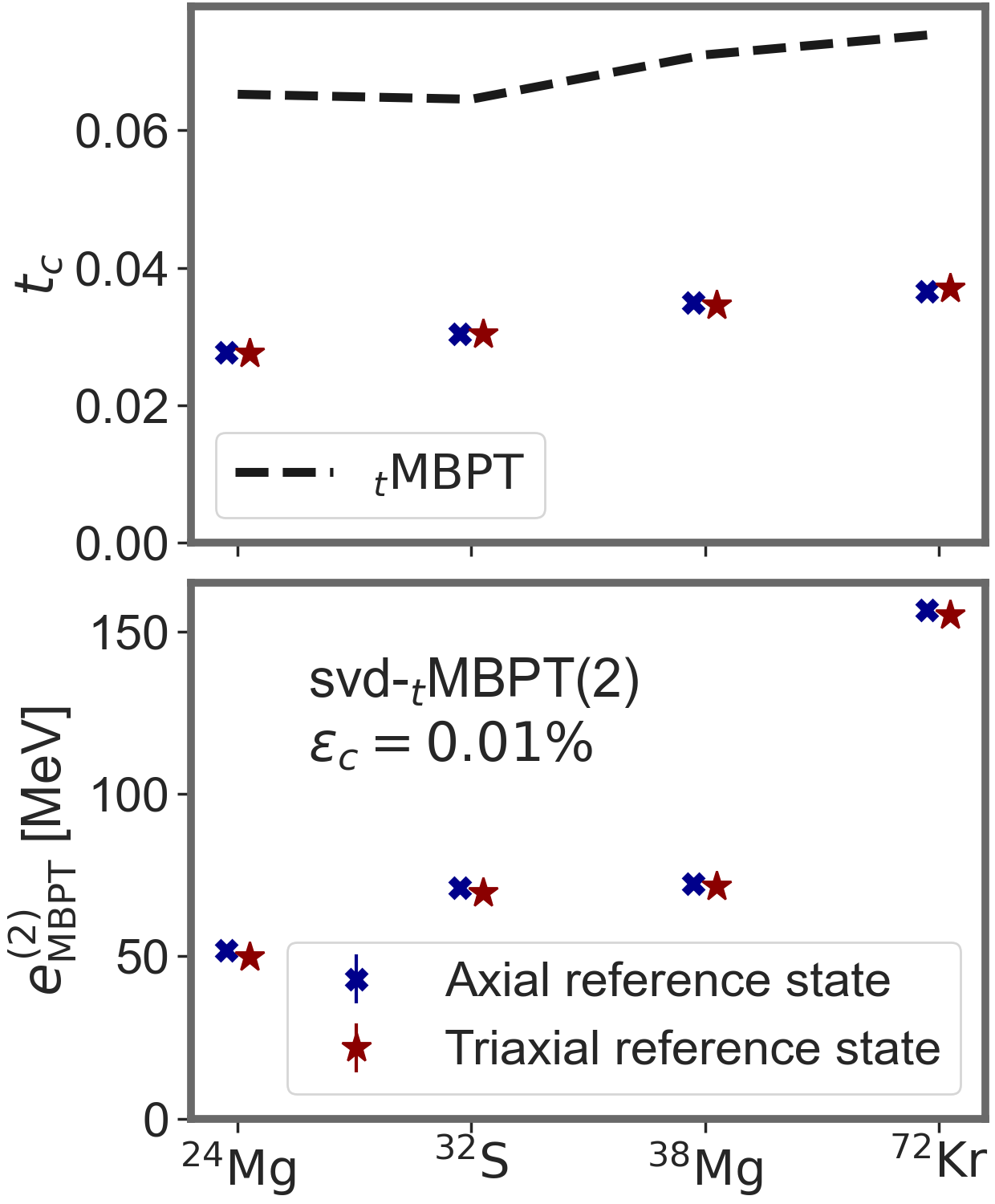}
    \caption{svd-$_\text{t}$MBPT(2) calculations of $^{24,38}$Mg, $^{32}$S and $^{72}$Kr performed for \(e_\text{max}=14\) based on prolate and triaxial reference states and employing the tolerated error \(\epsilon_\text{c}=0.01\%\). Top panel: \label{fig:tc}\label{fig:corr_tri} compression factor \(t_\text{c}\).  The dashed line corresponds to the compression factor associated with the optimal storage scheme in $_\text{t}$MBPT(2). Bottom panel: Second-order correlation energy $e^{(2)}_{\text{MBPT}}$, error bars correspond to a $0.01\%$ error stemming from the choice of \(\epsilon_{\text c}\) \correction{}{even if they are not visible on the scale of the figure.}}
\end{figure}

The upper panel of Figure~\ref{fig:tc} displays the compression factor \(t_\text{c}\) achieved in svd-$_\text{t}$MBPT(2)  with respect to the naive storage scheme on top of axial and triaxial reference states in the four nuclei of interest. The compression factor associated with the optimal storage scheme in $_\text{t}$MBPT(2) is also shown. The compression achieved via the central RSVD happens to be insensitive to the symmetry of the underlying reference state, \textit{i.e.} the irrelevant information compressed through the TTF does not relate to the possible axial symmetry of the reference state. To benefit from the latter, the associated sparsity of the resulting singular vectors must be further exploited explicitly, which requires to switch to the svd-$_\text{a}$MBPT(2) setting. 

The correlation energy $e^{(2)}_{\text{MBPT}}$ computed on top of the prolate and triaxial minima are compared in the lower panel of Fig.~\ref{fig:corr_tri}. With  the EM1.8/2.0 Hamiltonian, $e^{(2)}_{\text{MBPT}}$ ranges from $50$\,MeV in $^{24}$Mg to $150$\,MeV in  $^{72}$Kr. Albeit difficult to see on this scale, the correlation energy computed on top of the triaxial minimum is systematically smaller than for the prolate reference state. 

\begin{figure}
    \centering
    \includegraphics[width=.5\textwidth]{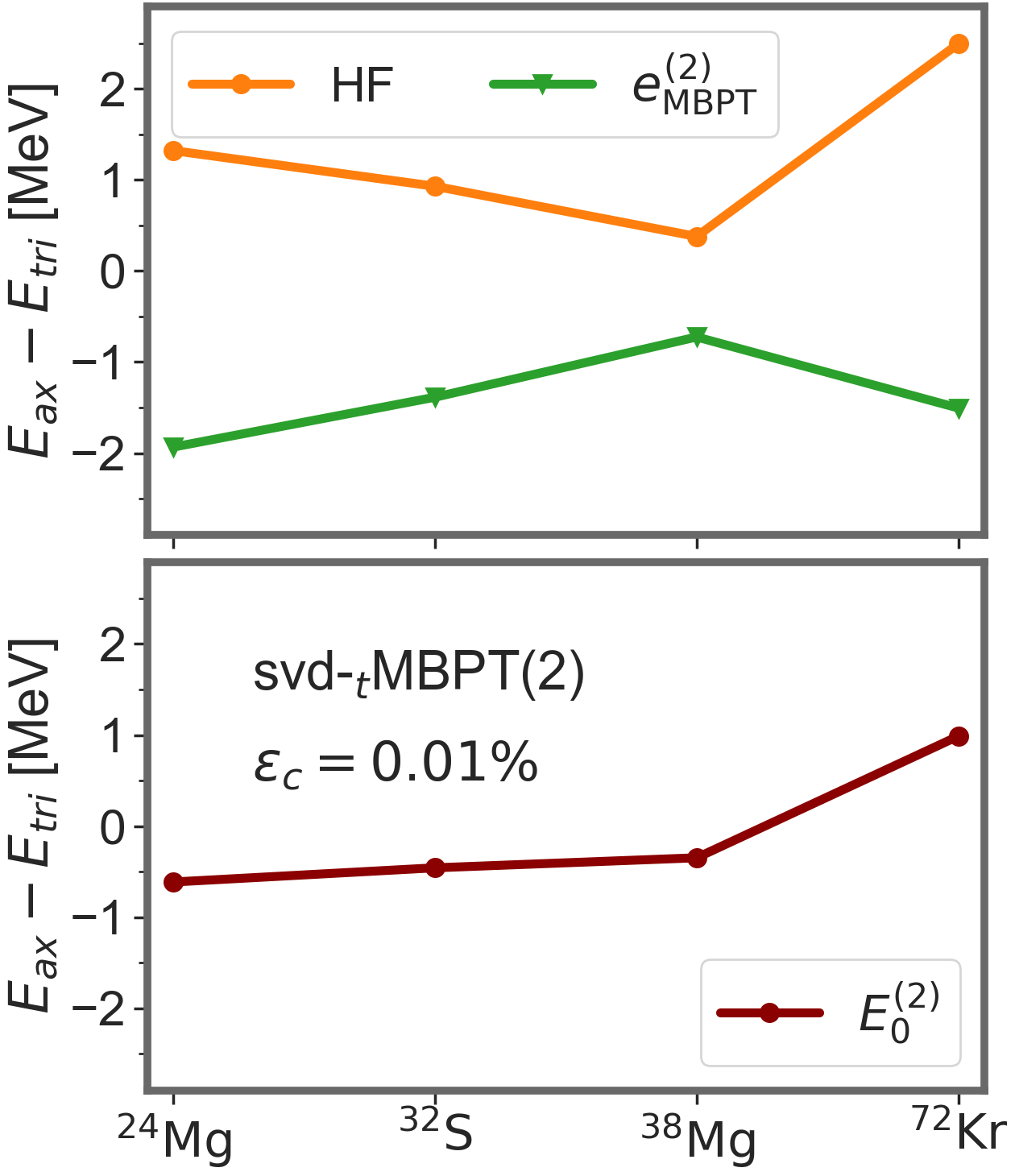}
    \caption{\label{fig:diff} Energy differences  computed via svd-$_\text{t}$MBPT(2) for a prolate reference state and for a  triaxial reference state.  Calculations are performed for \(e_\text{max}=14; \epsilon_\text{c}=0.01\%\). Upper panel: HF energy and second-order correction. Lower panel: total ground-state energy.}
\end{figure}

This deficit of correlation energy for the triaxial reference state compared to the prolate one is contrasted in the upper panel of Fig.~\ref{fig:diff} against the advantage obtained at the HF level. Because both differences are of similar size but opposite sign, the inclusion of many-body correlations flattens the total energy surface~\cite{Frosini22c} over the  \((\beta,\gamma)\) deformation plane compared to the HF level (Fig.~\ref{fig:hfb-pes}). This flattening obtained at MBPT(2) level is a sign of convergence of the perturbative expansion\footnote{In the exact limit the total energy surface would be exactly flat, \textit{i.e.} under the hypothesis that the many-body expansion under use converges, the exact ground-state energy is indeed independent of the nature, \textit{e.g.} intrinsic deformation, of the unperturbed reference state.}. Still, a non-zero difference remains at the MBPT(2) level as visible from the lower panel of Fig.~\ref{fig:diff}. Such a difference is of the order of a few hundreds of keVs. 

Eventually, while exploiting explicitly the triaxial degree of freedom seems to be unnecessary for absolute binding energies, it may be of interest to compute differential quantities while remaining at the MBPT(2) level, \textit{e.g.} two-neutron separation energies along  a series of nuclei with varying intrinsic deformations. For such a computation to be useful, a $0.01\%$ error on $e^{(2)}_{\text{MBPT}}$ (\textit{e.g.} $15$\,keV in $^{72}$Kr) is required, which is indeed presently achieved in large bases via svd-$_\text{t}$MBPT(2).

\subsection{BMBPT(2)}

As discussed in connection with Fig.~\ref{fig:scaling}, the dimensions at play in BMBPT are much larger than in MBPT such that the benefit of the TF is expected to be larger.  In the present section, calculations are performed on top of the paired oblate minimum in \nucl{Mg}{24}, \nucl{Mg}{38} and \nucl{Kr}{72} based on \(e_\text{max}=14\). 

\begin{figure}
    \centering\includegraphics[width=.5\textwidth]{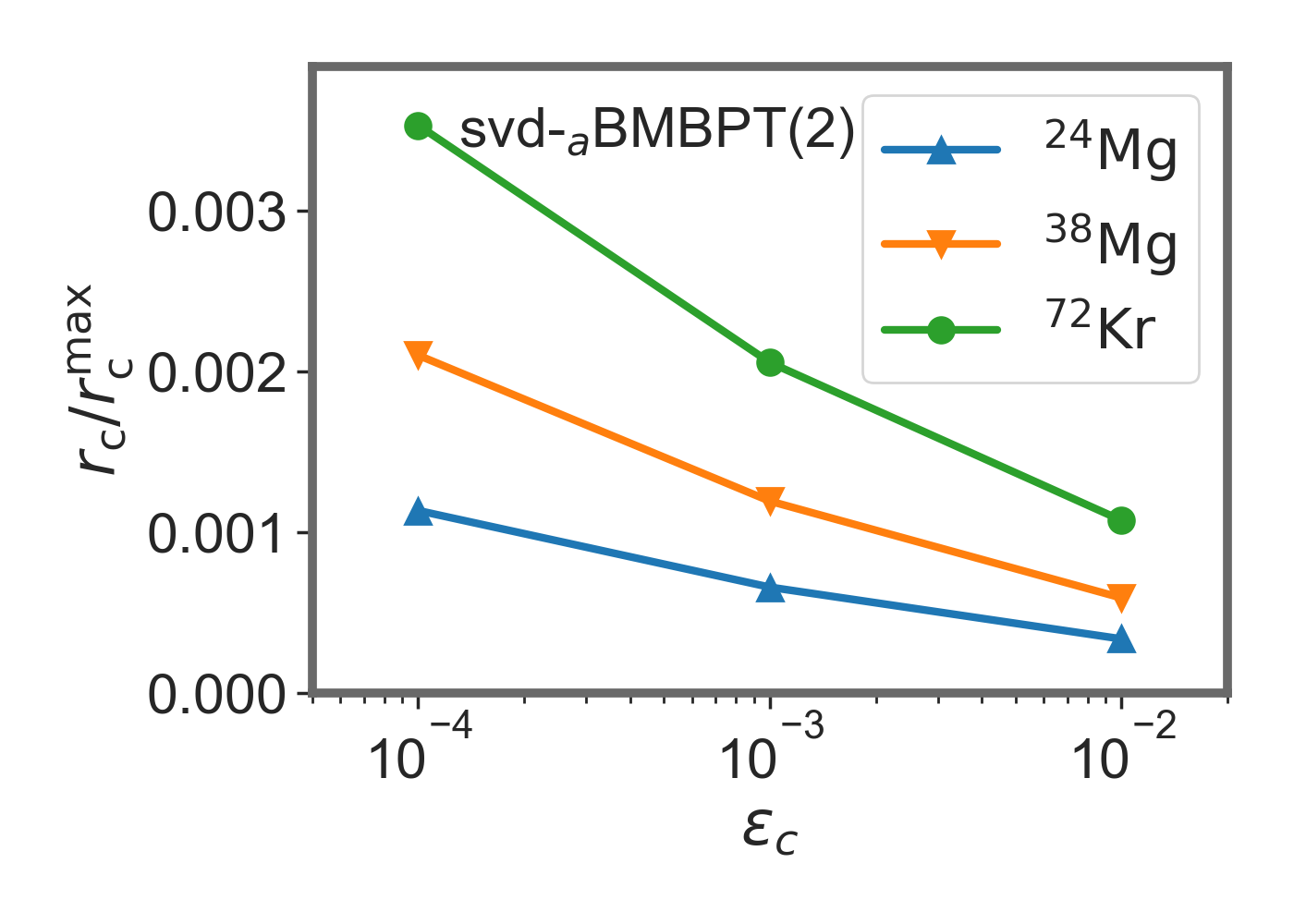}
    \caption{\label{fig:r_bmbpt} Ratio between central rank \(r_\text{c}\) and theoretical upper bound \(r_\text{c}^\text{max}=N^2=2720^2\) as a function of the RSVD truncation error \(\epsilon_\text{c}\) in svd-$_\text{a}$BMBPT(2) calculations for \nucl{Mg}{24}, \nucl{Mg}{38} and \nucl{Kr}{72}. Calculations are performed for \(e_\text{max}=14\).}
\end{figure}

Employing svd-$_\text{a}$BMBPT(2), the ratio of \(r_\text{c}\) to its theoretical upper bound \(r_\text{c}^\text{max}=N^2\) is shown in Fig.~\ref{fig:r_bmbpt} to decrease with the RSVD tolerance \(\epsilon_\text{c}\) and is typically two orders of magnitudes smaller than in svd-$_\text{a}$MBPT(2).  Similar results are obtained with svd$^2$-$_\text{t}$BMBPT(2). However, the ratio obtained for a given error is now nucleus dependent, \textit{i.e.} whereas the variation of the number of relevant degrees of freedom with $A$ was explicitly accounted for by the theoretical upper bound in MBPT, and thus canceled out in the ratio, it is not the case here where the upper bound is nucleus independent\footnote{In fact, it is observed (but not shown here) that when comparing \(r_{\text c}\) to \(A(N-A)\) instead of \(N^2\), the svd-MBPT behavior shown in Fig.~\ref{fig:r-mbpt} is recovered, which shows that BMBPT does not add as many relevant degrees of freedom as could be expected from the dimension analysis.}.

\begin{figure}
    \centering
    \includegraphics[width=.52\textwidth]{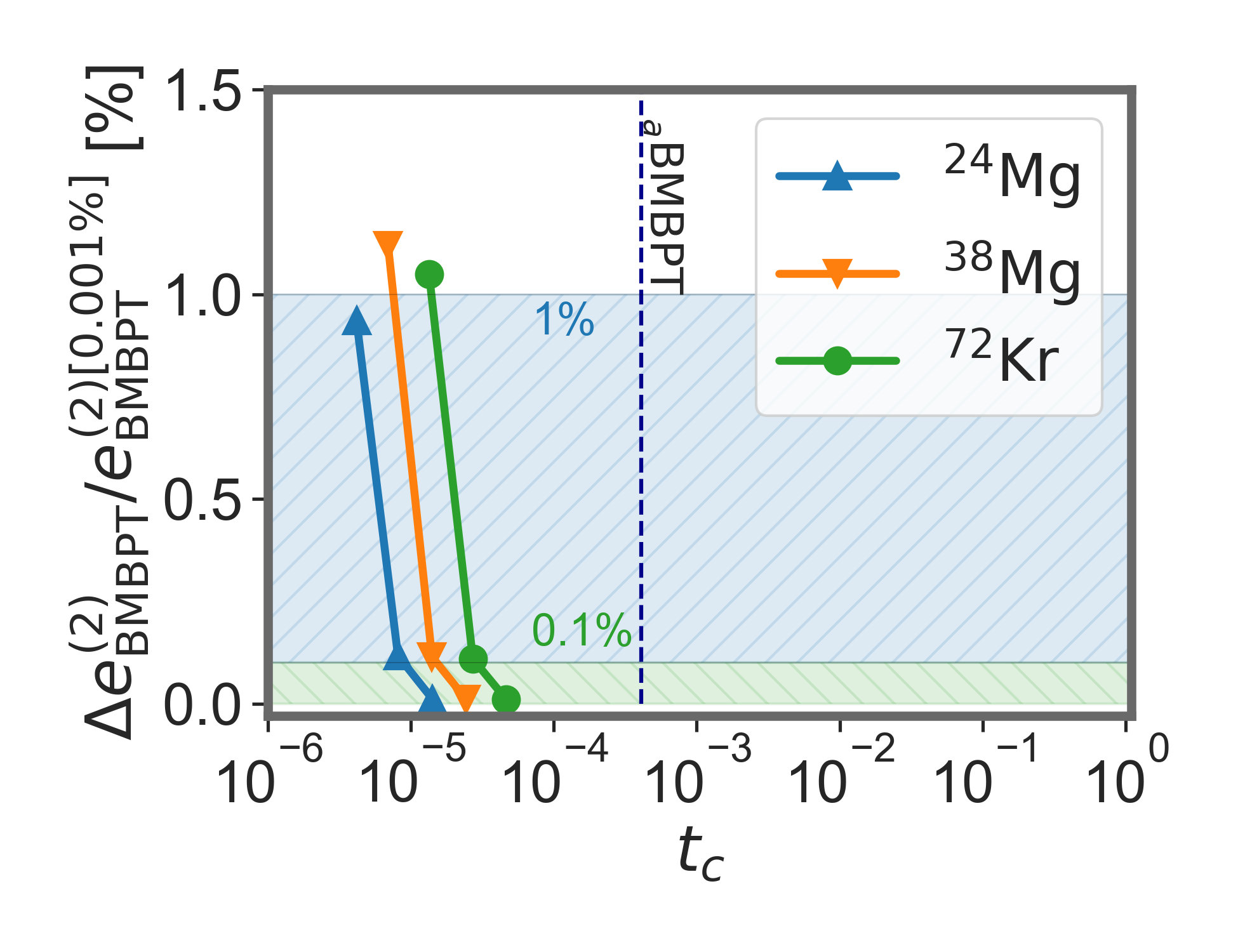}
    \caption{\label{fig:corr_btsvds} Error on $e^{(2)}_{\text{BMBPT}}$ based on paired oblate reference states in svd-$_\text{a}$BMBPT(2) calculations of \nucl{Mg}{24}, \nucl{Mg}{38} and \nucl{Kr}{72} against the compression factor \(t_\text{c}\) computed with respect to the naive storage scheme of $H^{40}_{k_1k_2k_3k_4}$. Points correspond to three values of the RSVD tolerance \(\epsilon_\text{c} \in \{1\%, 0.1\%, 0.01\%\}\) knowing that the error is computed with respect to the value obtained  for \(\epsilon_\text{c}=0.001\%\). The compression factor corresponding to the optimal storage scheme in $_\text{a}$BMBPT(2) is shown as a vertical blue dashed line. Regions corresponding to 0.1\% and 1\% error are highlighted. Calculations are performed for \(e_\text{max}=14\).}
\end{figure}

The error on the correlation energy $e^{(2)}_{\text{BMBPT}}$ obtained via svd-$_\text{a}$BMBPT(2)\footnote{The results obtained via svd$^2$-$_\text{t}$BMBPT(2), not shown here for simplicity, differ from those shown in Fig.~\ref{fig:corr_btsvds} by less than $0.2\%$ while reaching a compression factor that is only a factor of 2 larger. The latter result is possible thanks to the second RSVD employed in svd$^2$-$_\text{t}$BMBPT(2) that is able to  achieve a similar compression as the explicit sparse format exploited in svd-$_\text{a}$BMBPT(2). This shows the potential of svd$^2$-$_\text{t}$BMBPT(2) whenever a paired triaxial reference state needs to be employed.} is displayed in Fig~\ref{fig:corr_btsvds} as a function of \(t_\text{c}\) for three tolerated errors \(\epsilon_\text{c}\in\{0.01\%,0.1\%,1\%\}\)  in \nucl{Mg}{24}, \nucl{Mg}{38} and \nucl{Kr}{72}. As the reference value is not available for deformed BMBPT(2) in $e_\text{max}=14$, the error is computed with respect to the value obtained for \(\epsilon_\text{c}= 0.001\%\). The results demonstrate that the conclusions reached for MBPT(2) calculations are also valid in BMBPT(2) while achieving greater compression factors.

\subsection{Discussion and perspectives}

\correction{
The full benefit of TTF is expected to arise in many-body calculations requiring (i) very large bases and (ii) mode-6 many-body tensors (iii) of which several instances must be stored. 
\begin{figure}
    \centering
    \includegraphics[width=.52\textwidth]{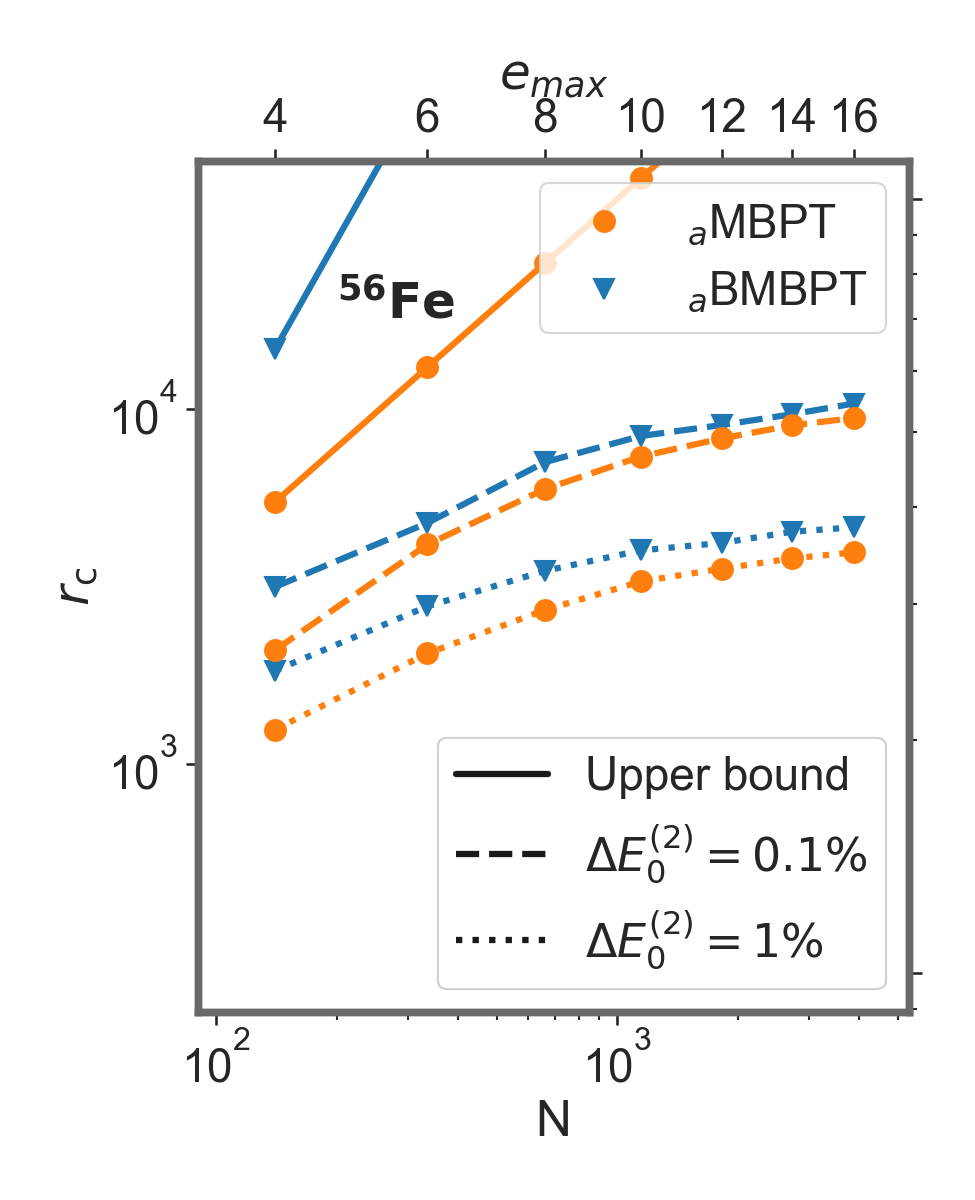}
    \caption{\label{fig:fe}  Rank \(r_\text{c}\) as a function of \(N\) in svd-$_\text{a}$MBPT(2) and svd-$_\text{a}$BMBPT(2) calculations of  \nucl{Fe}{56}. Results are displayed for two values $\Delta E^{(2)}_0 \in \{1\%,0.1\%\}$ of the error on the total ground-state energy. Two lines show the corresponding theoretical upper bound corresponding to keeping all eigenvalues.}
\end{figure}

The present work is making a significant step forward regarding point (i), both by actually considering large initial basis sizes and by employing expansion methods based on (triaxially) deformed and potentially superfluid unperturbed states. To further illustrate the potentiality of TTF in this context, svd-(B)MBPT results are now characterized as a function of the basis size dimension  with \(e_\text{max}\) ($N$) ranging from 4 to 16 (140 to 3876).
}{Truncated TF is expected to be more beneficial as the size of many-body tensors increases, allowing to push to higher bases and/or to higher rank operators.
\begin{figure}
    \centering
    \includegraphics[width=.52\textwidth]{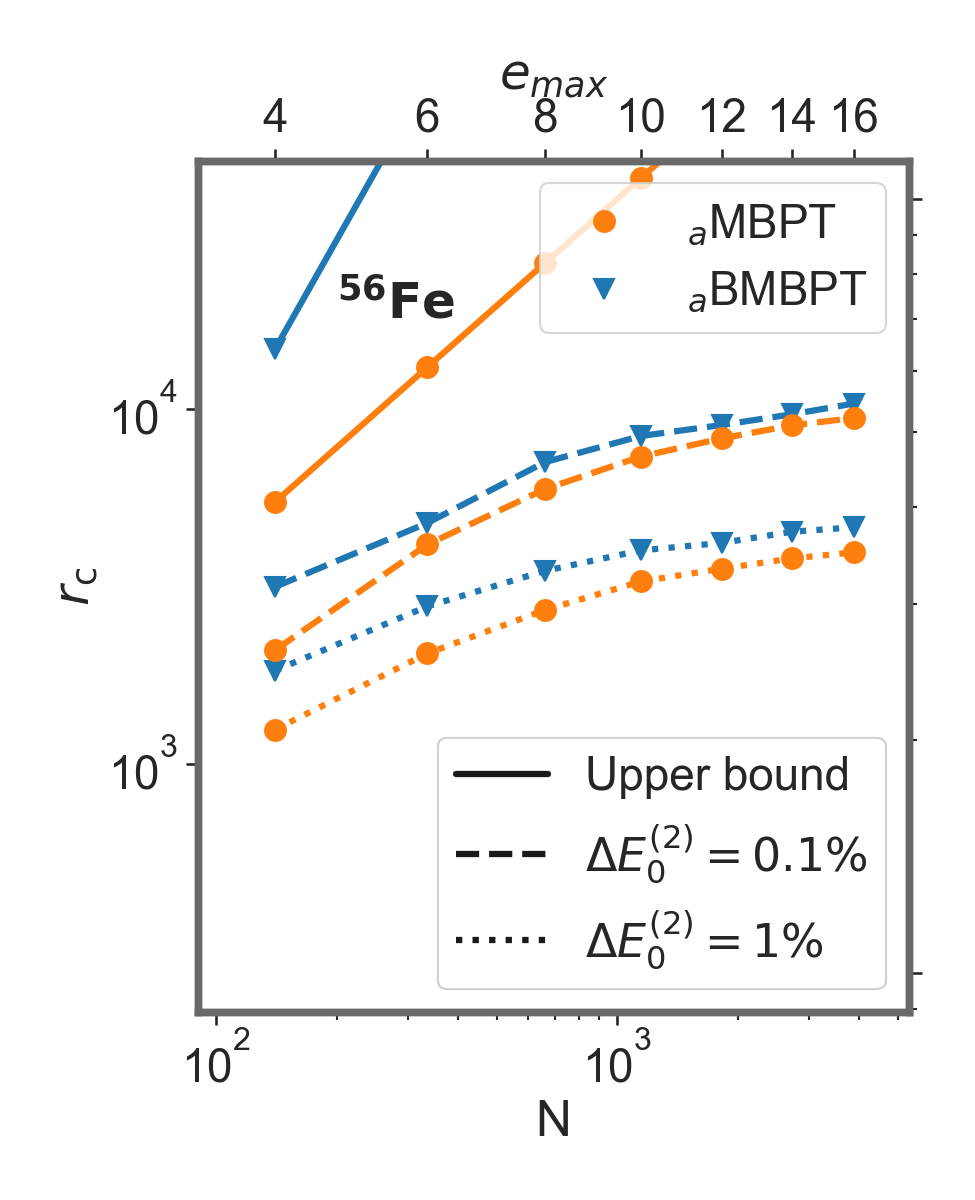}
    \caption{\label{fig:fe}  Rank \(r_\text{c}\) as a function of \(N\) in svd-$_\text{a}$MBPT(2) and svd-$_\text{a}$BMBPT(2) calculations of  \nucl{Fe}{56}. Results are displayed for two values $\Delta E^{(2)}_0 \in \{1\%,0.1\%\}$ of the error on the total ground-state energy. Two lines show the corresponding theoretical upper bound corresponding to keeping all eigenvalues.}
\end{figure}
This work allows in particular to enlarge the size of the one-body basis and employ expansion methods breaking many symmetries. To further illustrate the potentiality of TTF in this context, svd-(B)MBPT results are now characterized as a function of the basis size dimension  with \(e_\text{max}\) ($N$) ranging from 4 to 16 (140 to 3876).}
 
Figure~\ref{fig:fe} first displays the rank \(r_\text{c}\) of the central RSVD as a function of $N$ in \nucl{Fe}{56}. The  svd-$_\text{a}$MBPT(2)  (svd-$_\text{a}$BMBPT(2)) calculations employ the unpaired prolate (paired oblate) reference state and results are shown for an error\footnote{This error is controlled via appropriate choices of \(\epsilon_\text{c}\).} $\Delta E^{(2)}_0 \in \{1\%,0.1\%\}$ on the {\it total} ground-state energy\footnote{For a low resolution-scale nuclear Hamiltonian as the one presently employed, $1\%$ ($0.1\%$) of the total second-order ground-state energy $E^{(2)}_0$ is of the order (of one tenth) of the next order correction $e^{(3)}_{\text{(B)MBPT}}$. Thus, requiring $1\%$ error from the TF is enough to perform accurate (B)MBPT(2) calculations whereas requesting $0.1\%$ error constitutes the target accuracy for going safely to the next truncation order~\cite{Scalesi24}.}. While the theoretical upper bound \(r^{\text{max}}_\text{c}\) is proportional to \(N\) in MBPT(2) and to  \(N^2\) in BMBPT(2), \(r_\text{c}\) grows at a much slower rate.  Regardless of \(\epsilon_\text{c}\), \(r_\text{c}\) flattens out asymptotically, underlining the optimal benefit of the TTF  in very large bases.

\begin{figure*}
    \centering
    \includegraphics[width=1.00\textwidth]{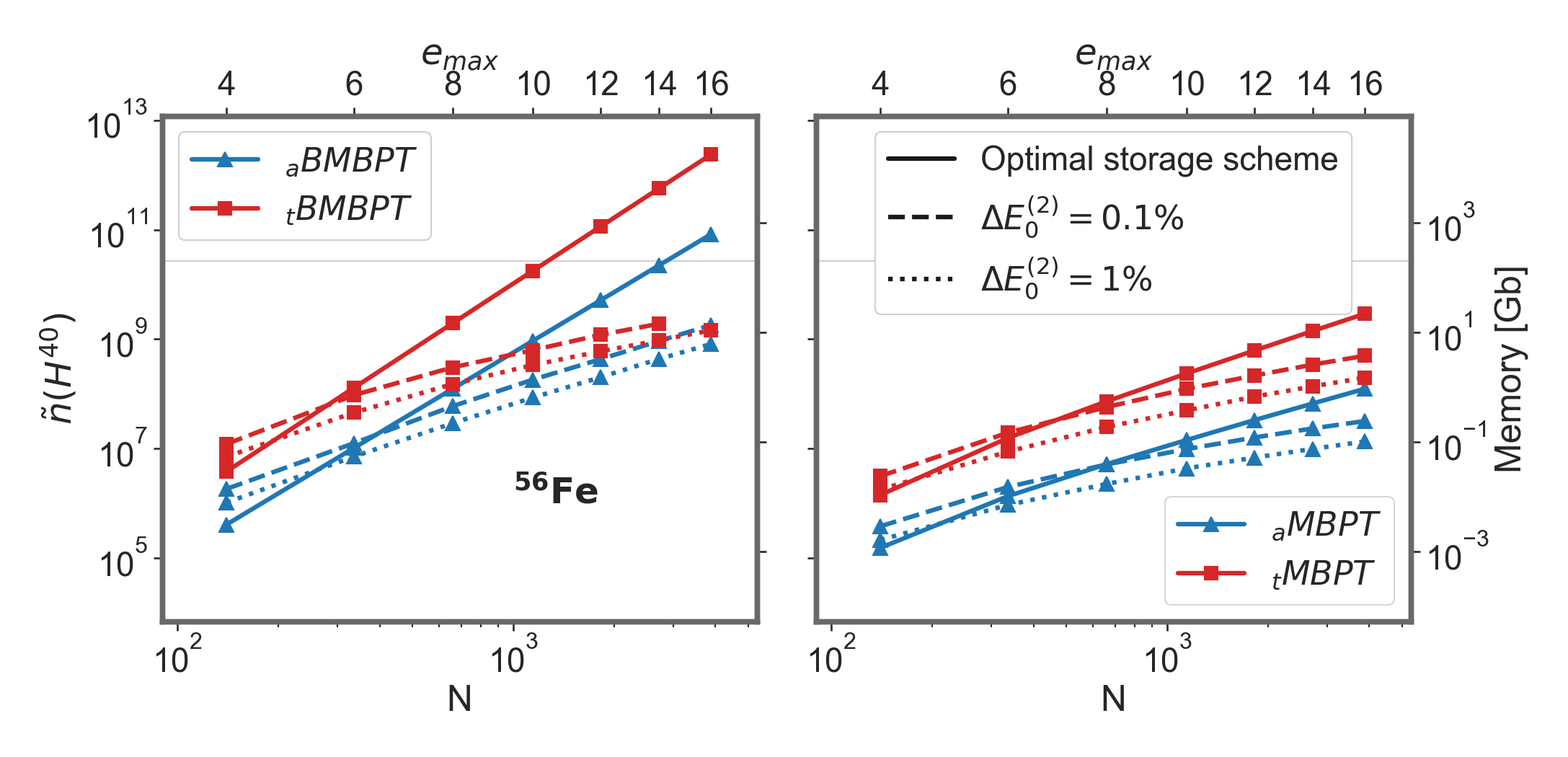}
    \caption{\label{fig:fecompression} Storage requirement as a function of \(N\) in TF calculations of  \nucl{Fe}{56} compared to the optimal storage schemes in the exact (B)MBPT(2) calculations. Results are displayed for two values $\Delta E^{(2)}_0 \in \{1\%,0.1\%\}$ of the error on the total ground-state energy. Left panel: svd-$_\text{a}$BMBPT(2) and svd$^2$-$_\text{t}$BMBPT(2) calculations based on the paired oblate reference state. Right panel: svd-$_\text{a}$MBPT(2) and svd-$_\text{t}$MBPT(2) calculations based on the unpaired prolate reference state.}
\end{figure*}

Furthermore, the comparison of svd-$_\text{a}$MBPT(2) and svd-$_\text{a}$BMBPT(2) results reveal that the central rank is not very sensitive to the presence of pairing correlations. This is particularly true asymptotically where both sets of curves become close to one another. 
Such a behavior reflects the fact that pairing correlations are of infrared character and, as such, limited to a certain window around the Fermi energy. Hence, such correlations do not effectively add a large number of relevant degrees of freedom. While expansion methods based on Bogoliubov reference states such as BMBPT and BCC~\cite{Arthuis:2018yoo,SiDu15}, as well as Gorkov self-consistent Green's function~\cite{SoDu11} and Bogoliubov in-medium similarity renormalization group~\cite{tichai2021adg}, do overcomplexify the problem to grasp such strong static correlations\footnote{It is key to note that such an overcomplexification is still very limited compared to methods attempting to capture such static correlations without breaking U(1) symmetry. The same is true regarding the incorporation of strong quadrupole correlations via the breaking of rotational symmetry.}, TF naturally filters out the redundant degrees of freedom as presently exemplified. 

In the right (left) panel of Fig.~\ref{fig:fecompression}, the storage requirement in svd-$_\text{a}$(B)MBPT(2) and svd-$_\text{t}$(B)MBPT(2) calculations of  \nucl{Fe}{56} are compared to  the optimal storage schemes in exact (B)MBPT(2) calculations. Results are displayed for the same two errors $\Delta E^{(2)}_0 \in \{1\%,0.1\%\}$ as in Fig.~\ref{fig:fe}. In very small bases, the TF does not bring any gain due to the fact that antisymmetry cannot be (can only be partially) exploited in the TTF employed for (B)MBPT calculations. As the basis size increases, the benefit from the TTF becomes more and more important compared to the optimal storage scheme of  unfactorized calculations. 

In MBPT(2), the storage is eventually reduced by one order of magnitude at  \(e_\text{max}=16\) for $\Delta E^{(2)}_0 = 1\%$, effectively corresponding to a \(e_\text{max}\approx 8-10\) calculation in the original setting. 

In BMBPT(2), the gain is much greater compared to the original cost that is itself much larger. At \(e_\text{max}=16\) the gain is typically two orders of magnitude for axial calculations and four orders of magnitude for triaxial ones\footnote{Due to the specificity of the computer architecture used to perform the calculations, some further optimization would be needed to perform  the svd-$_\text{t}$BMBPT(2) calculations in \(e_\text{max}=16\) with $\Delta E^{(2)}_0 = 0.1\%$ such that the corresponding point is missing. It is not a problem for the present discussion given that the extrapolation from the available points is trivial to perform.}, corresponding respectively to an effective \(e_\text{max}\approx 11\) and \(e_\text{max}\approx 7\)  calculation in the original setting. Interestingly, the cost of svd-$_\text{t}$BMBPT(2) is eventually not much greater than the cost of svd-$_\text{t}$MBPT(2).

While going to $\Delta E^{(2)}_0 = 0.1\%$ reduces the gain on the storage of $H^{40}$, in particular for MBPT(2), it is only needed in cases where the storage cost will anyway be dominated by other many-body tensors for which the relative benefit will be much greater, thus maintaining the relevance of the TTF.

\section{Conclusions}
\label{secconclusions}

Tensor factorization techniques are expected to be beneficial to overcome the storage bottleneck arising in \textit{ab initio} many-body calculations requiring (i) very large single bases and (ii) mode-6, \textit{i.e.} three-body, tensors (iii) that must be stored repeatedly. In this context, the present work achieved a significant step forward regarding point (i) by indeed considering large one-body basis sizes and by employing expansion methods based on (triaxially) deformed and potentially superfluid unperturbed states, which forbids the use of symmetry consideration to effectively reduce the initial basis dimension. 

The use of randomized projection techniques combined with matrix-free products was key to achieve this goal, \textit{i.e.} the heavy tensor at play in deformed (B)MBPT could be factorized without ever needing to compute (and store) it explicitly.  This allowed us to perform a first application in systems that spontaneously break axial symmetry at the mean-field level in nuclei as heavy as $^{72}$Kr.

While the benefit is already significant in the present setting, \textit{e.g.} allowing calculations that cannot be performed without TF, it is now necessary to move to steps (ii) and (iii) in order to achieve much greater benefits. The first objective is to extend the present work to (B)MBPT(3) calculations of deformed nuclei in order to tackle a formalism that becomes rapidly limited by memory usage. In a second stage, the formats presented here could be employed as appropriate ansätze to factorize non-perturbative formalisms such as deformed (B)CC or (B)IMSRG.

\begin{acknowledgements}

The authors would like to thank Alberto Scalesi and Heiko Hergert for generating the interaction files as well as Lars Zurek for proofreading the paper. The work of M.F. and P.T. was supported by the CEA-SINET project. This work was performed using CCRT HPC resource (TOPAZE supercomputer) and using HPC resources from GENCI - [TGCC/CINES/IDRIS] (Grant 2023- [A0150513012]).

\end{acknowledgements}

\begin{appendix}

\section{Randomized SVD}\label{RandSVD}

    In this section, a matrix \({\correction{A}{W}}\in\mathbb R^{m\times n}\) is considered. The goal is to compute a rank-$r$ truncated SVD decomposition of \({\correction{A}{W}}\), \textit{i.e.}
    \begin{equation}
        {\correction{A}{W}}\approx \tilde {\correction{A}{W}} = C s D,
    \end{equation}
    where \(C\in \mathbb R^{m\times r}, D\in \mathbb R^{r\times n}, s\in \mathbb R^{r\times r}\), with \(s\) a diagonal matrix.

    \subsection{RSVD algorithm}

        \subsubsection{General case}

    Given an approximate orthonormal basis \(Q\) of the range of \({\correction{A}{W}}\) with dimensions \( m \times r\), the RSVD method computes a rank-$r$ approximation of \({\correction{A}{W}}\) following~\cite{martinsson2021randomized}
    \begin{equation}
        \tilde {\correction{A}{W}}= Q Q^\transpose {\correction{A}{W}}.
    \end{equation}
    The quality of the approximation, monitored by the Frobenius norm of the difference
    \begin{equation}
        \epsilon \equiv \frac{{\|{\correction{A}{W}}-\tilde{{\correction{A}{W}}}\|}_\text{F}}{\|{\correction{A}{W}}\|_\text{F}},
    \end{equation}
    is therefore directly dependent on the quality of \(Q\). 
    The construction of \(Q\) via a randomized range-finder algorithm is detailed in the next subsection.

    The SVD decomposition of \(\tilde {\correction{A}{W}}\) is obtained from the SVD decomposition of \(Q^\transpose {\correction{A}{W}}\) as
        \begin{align}
    \nonumber        \tilde {\correction{A}{W}} &= QQ^\transpose {\correction{A}{W}} &&\\
    \nonumber        &= Q (Q^\transpose {\correction{A}{W}})& & \\
    \nonumber        &\equiv Q (\hat C s D) & &\text{SVD on } Q^\transpose {\correction{A}{W}}\\
    \nonumber        &\equiv C s D. &&\text{Defining } C\equiv Q\hat C
        \end{align}

    \subsubsection{Symmetric case}
    \label{appsymdiago}

Whenever \(A\in \mathbb R^{n\times n}\) is symmetric, the previous algorithm can be replaced with a truncated symmetric eigenvalue decomposition such that only left eigenvectors are stored. Indeed, defining a symmetric approximation of \({\correction{A}{W}}\) as
    \begin{equation}
        \tilde {\correction{A}{W}}\equiv QQ^\transpose {\correction{A}{W}}QQ^\transpose.
    \end{equation}
     \(Q^\transpose {\correction{A}{W}}Q\) is still a real symmetric matrix that can be decomposed according to
    \begin{equation}
        Q^\transpose {\correction{A}{W}}Q \equiv \hat C s \hat C^\transpose.
    \end{equation}
    Multiplying on the left by \(Q\) gives vectors \(C\equiv Q \hat C\) in the original basis such as
    \begin{equation}
        \tilde {\correction{A}{W}} = C s C^\transpose.
    \end{equation}

    \subsection{Adaptative Lanczos range-finder}
    \label{adaptrange-finder}

The determination of \(Q\) is the key to build an accurate low-rank approximation of \({\correction{A}{W}}\). 
    Several algorithms exist~\cite{martinsson2021randomized} to construct an optimal \(Q\) from a set of randomly generated vectors. 
    To do so, it is essential that singular vectors corresponding to the largest singular values of \({\correction{A}{W}}\) are well reproduced by \(Q\). 
    The method chosen in this work is the adaptative block-Lanczos range-finder. 
    This algorithm constructs \(Q\) in the Krylov space \(\{{\correction{A}{W}}X,{\correction{A}{W}} {\correction{A}{W}}^\transpose {\correction{A}{W}}X,\cdots,(AA^\transpose)^q{\correction{A}{W}}X\}\) of a given set of randomly chosen vectors \(X\in\mathbb R^{n\times b}\). 
    
    The depth \(q\) of the Krylov space offers a natural way to iteratively increase the search range without adding much complexity. 
    This is mainly useful for practical application if there exists a way to assess the error \(\epsilon\) associated with any given \(Q\). Originally, the Lanczos algorithm was designed with \(b=1\) which requires very large \(q\). For range-finder applications there is no {\it a priori} way to determine the optimal \(b\). In this work we chose to use \(b\) of the order of 100 and adapt \(q\) consistently.

    \subsubsection{Error estimator\label{sec:StochEstim}}

    It is impossible to evaluate exactly the error in practice for large matrices given that computing \(\|{\correction{A}{W}}\|_\text{F}\) and \(\|(1-QQ^\transpose) {\correction{A}{W}}\|_\text{F}\) requires the matrix elements of \({\correction{A}{W}}\). Reference~\cite{tropp2023randomized} proposes to replace the computation of the exact norm by stochastic estimators that are inexpensive to compute but accurate enough. 
    The key is to introduce a small training set \(X_{\text{test}}\sim\mathcal N(0,1_{n\times l})\), where \(l\approx 20\) is very small compared to the original dimensions. Defining \(Z_{\text{test}}\equiv {\correction{A}{W}}X_{\text{test}}\),
    the exact norms at play are replaced by the low-dimensional estimators
    \begin{subequations}
        \begin{align}
            \|{\correction{A}{W}}\|_\text{F}^2 &\approx \frac 1l \|Z_{\text{test}}\|^2_\text{F},\\
            \|(1-QQ^\transpose){\correction{A}{W}}\|_\text{F}^2 &\approx \frac 1l \|(1-QQ^\transpose)Z_{\text{test}}\|^2_\text{F},
        \end{align}
    \end{subequations}
    such that the estimated error is given by
    \begin{equation}\label{eq:stoc_est}
        \epsilon_{\text{c}} \equiv \frac{\|(1-QQ^\transpose)Z_{\text{test}}\|_\text{F}}{\|Z_{\text{test}}\|_\text{F}}.
    \end{equation}
    Numerical tests have shown that the estimated relative error is very reliable in practice.

    \subsubsection{Algorithm}

    Details on the Lanczos range-finder are given in Alg.~\ref{alg:lanczos-rf}, following~\cite{martinsson2021randomized}.
    \begin{algorithm}
\caption{Lanczos range-finder\label{alg:lanczos-rf}}
    \KwData{\({\correction{A}{W}}\in \mathbb R^{m\times n}, \eta\in \mathbb R^+, b\in\mathbb N, l\in \mathbb N\)}

\tcc{Draw testing set used for error estimation}

    Draw \(X_\text{test} \sim N(0, 1_{n\times l})\)

    \(Z_\text{test}\gets {\correction{A}{W}}X_\text{test}\);

    \(n_\text{ref}\gets \|Z_\text{test}\|_\text{F};\ \epsilon_\text{c}\gets1\);

    \tcc{Initial vector seeds for Krylov space}

    Draw \(X \sim N(0, 1_{n\times b})\);
    
    \(Z\gets AX\);

    \([Q_0,] = \texttt{qr\_econ}(Z)  \)  ;
    
    \(W = {\correction{A}{W}}^\transpose Q_0 \)  ;
    
    \([P_0,R] = \texttt{qr\_econ}(W)  \)  ;

    \(i\gets1\)
    
        \tcc{Lanczos procedure}
    \KwRepeat{ \texttt{true} }{ 
        \tcc{Error estimation}
        \KwFor{$j=0\cdots i-1$}
        {
        \tcc{Orthogonalisation}
        \texttt{stable\_\correction{gram\_schmidt}{orth}}$(Z_\text{test},[Q_0,\cdots,Q_{i-1}])$ 
        }
        \(\epsilon_{\text c} \gets \frac{\|Z_\text{test}\|_\text{F}}{n_\text{ref}}\);
        
        \lIf{$\epsilon_\text{c}<\eta$}{\rm{\bf{break}}}

        \tcc{Lanczos recursion}
        $Z = {\correction{A}{W}} P_{i-1} - Q_{i-1} R^\transpose$;

        \texttt{stable\_\correction{gram\_schmidt}{orth}}$(Z,[Q_0,\cdots,Q_{i-1}])$;

        \([Q_i,R]=\texttt{qr\_econ}(Z)\);

        \(W = {\correction{A}{W}}^\transpose Q_i - P_{i-1} R^\transpose\);

        \texttt{stable\_\correction{gram\_schmidt}{orth}}$(W,[P_0,\cdots,P_{i-1}])$;

        \([P_i,R]=\texttt{qr\_econ}(W)\);

        $i\gets i+1$;
    }

    \KwReturn{$Q=[Q_0,\cdots,Q_i]$}

\end{algorithm}

\subsection{Example\label{sec:examplerf}}

The benefits of performing RSVD based on the Lanczos range-finder for efficient matrix decompositions is now illustrated on a schematic example. Given a symmetric matrix \({\correction{A}{W}}\) of dimension \(N=3000\) constructed such that its eigenvalues are decaying quadratically (e. g. \(\lambda_i=(i+10)^{-2}\)), its low-rank approximation is built employing the previously introduced method. 

\begin{figure}
    \centering
    \includegraphics[width=.5\textwidth]{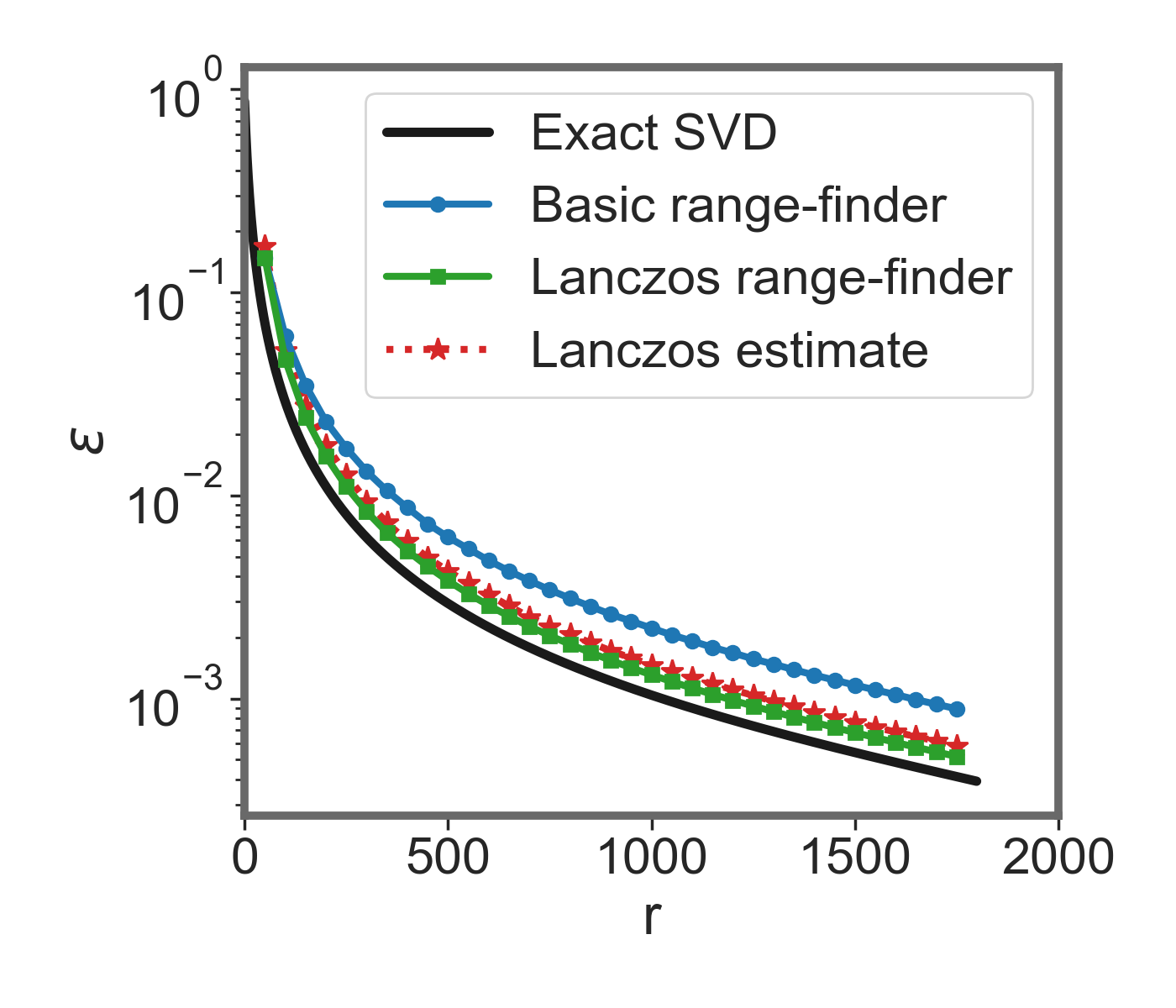}
    \caption{Error of low rank approximation of schematic symmetric matrix \({\correction{A}{W}}\) introduced in \ref{sec:examplerf} as a function of the rank \(r\) using exact SVD (optimal approximation), basic and Lanczos range-finder based RSVD respectively. A stochastic estimate of the error calculated at each Lanczos step is also displayed alongside the exact errors.}
    \label{fig:rfcomp}
\end{figure}

In order to exemplify the benefits of the adaptative Lanczos range-finder, it is compared to so called ``basic'' range-finder that is commonly used in most applications. In the latter case, a rank-$r$ approximation \(\tilde {\correction{A}{W}}(r)\) is obtained via the following steps
\begin{enumerate}
    \item Draw \(X\sim N(0,1_{n\times r})\),
    \item $Q \equiv \verb|orth|({\correction{A}{W}} X)$,
    \item \(\tilde {\correction{A}{W}}(r) \equiv QQ^T{\correction{A}{W}}\).
\end{enumerate}
In that case, \(r\) has to be chosen in advance and the error associated to the approximation is only calculated {\it a posteriori}. The Lanczos range-finder is used on the same matrix by choosing an initial set of \(b=50\) randomly generated vectors and by building successive rank-\(qb\) approximations of \({\correction{A}{W}}\) via the increase of the depth \(q\) following Alg~\ref{alg:lanczos-rf}.

Fig.~\ref{fig:rfcomp} displays the relative norm error \(\epsilon\) of the low-rank approximation of \({\correction{A}{W}}\) using the exact SVD decomposition, the basic range-finder and the Lanczos range-finder as a function of the rank of the approximation. It appears that the Lanczos range-finder, although not significantly more costly than the basic range-finder, provides an approximation of the original matrix that is much closer to the optimal one given by the exact SVD.

The main advantage of the Lanczos range-finder further lies in the possibility to make it adaptive by building a (inexpensive) stochastic estimate of the error (Eq.~\eqref{eq:stoc_est}). As seen in Fig.~\ref{fig:rfcomp}, such a stochastic estimate turns out to be very close to the deterministic evaluation. This validates its use as a stopping criterion to determine  the optimal value of \(q\) matching a required tolerance on the error in an adaptive fashion with a negligible overhead, as shown in Alg~\ref{alg:lanczos-rf}.

\end{appendix}

\bibliography{bibliography.bib}

\begin{thebibliography}{10}
\expandafter\ifx\csname url\endcsname\relax
  \def\url#1{\texttt{#1}}\fi
\expandafter\ifx\csname urlprefix\endcsname\relax\def\urlprefix{URL }\fi
\expandafter\ifx\csname href\endcsname\relax
  \def\href#1#2{#2} \def\path#1{#1}\fi

\bibitem{Hergert:2020bxy}
H.~Hergert, {A Guided Tour of $ab$ $initio$ Nuclear Many-Body Theory}, Front. in Phys. 8 (2020) 379.
\newblock \href {http://arxiv.org/abs/2008.05061} {\path{arXiv:2008.05061}}, \href {https://doi.org/10.3389/fphy.2020.00379} {\path{doi:10.3389/fphy.2020.00379}}.

\bibitem{Dickhoff2004}
W.~Dickhoff, C.~Barbieri, Progress in Particle and Nuclear Physics\href {https://doi.org/10.1016/J.PPNP.2004.02.038} {\path{doi:10.1016/J.PPNP.2004.02.038}}.

\bibitem{Hergert16a}
H.~Hergert, S.~Bogner, T.~Morris, A.~Schwenk, K.~Tsukiyama, {The in-medium similarity renormalization group: A novel ab initio method for nuclei}, Physics reports 621 (2016) 165--222.

\bibitem{Hagen2010}
G.~Hagen, T.~Papenbrock, D.~J. Dean, M.~Hjorth-Jensen, \href{https://link.aps.org/doi/10.1103/PhysRevC.82.034330}{{{\textless}i{\textgreater}Ab initio{\textless}/i{\textgreater} coupled-cluster approach to nuclear structure with modern nucleon-nucleon interactions}}, Physical Review C 82~(3) (2010) 034330.
\newblock \href {https://doi.org/10.1103/PhysRevC.82.034330} {\path{doi:10.1103/PhysRevC.82.034330}}.
\newline\urlprefix\url{https://link.aps.org/doi/10.1103/PhysRevC.82.034330}

\bibitem{tichai2020many}
A.~Tichai, R.~Roth, T.~Duguet, {Many-body perturbation theories for finite nuclei}, Front. in Phys. 8 (2020) 164.
\newblock \href {http://arxiv.org/abs/2001.10433} {\path{arXiv:2001.10433}}, \href {https://doi.org/10.3389/fphy.2020.00164} {\path{doi:10.3389/fphy.2020.00164}}.

\bibitem{Soma:2013xha}
V.~Som\`a, A.~Cipollone, C.~Barbieri, P.~Navr\'atil, T.~Duguet, {Chiral two- and three-nucleon forces along medium-mass isotope chains}, Phys. Rev. C 89~(6) (2014) 061301.
\newblock \href {http://arxiv.org/abs/1312.2068} {\path{arXiv:1312.2068}}, \href {https://doi.org/10.1103/PhysRevC.89.061301} {\path{doi:10.1103/PhysRevC.89.061301}}.

\bibitem{Tichai18BMBPT}
A.~Tichai, P.~Arthuis, T.~Duguet, H.~Hergert, V.~Som{\'a}, R.~Roth, {Bogoliubov many-body perturbation theory for open-shell nuclei}, {Physics Letters B} {786} ({2018}) {195--200}.

\bibitem{Yao:2019rck}
J.~Yao, B.~Bally, J.~Engel, R.~Wirth, T.~Rodr\'iguez, H.~Hergert, {$Ab$ $Initio$ Treatment of Collective Correlations and the Neutrinoless Double Beta Decay of $^{48}$Ca}, Phys. Rev. Lett. 124~(23) (2020) 232501.
\newblock \href {http://arxiv.org/abs/1908.05424} {\path{arXiv:1908.05424}}, \href {https://doi.org/10.1103/PhysRevLett.124.232501} {\path{doi:10.1103/PhysRevLett.124.232501}}.

\bibitem{Novario:2020kuf}
S.~J. Novario, G.~Hagen, G.~R. Jansen, T.~Papenbrock, {Charge radii of exotic neon and magnesium isotopes}, Phys. Rev. C 102~(5) (2020) 051303.
\newblock \href {http://arxiv.org/abs/2007.06684} {\path{arXiv:2007.06684}}, \href {https://doi.org/10.1103/PhysRevC.102.051303} {\path{doi:10.1103/PhysRevC.102.051303}}.

\bibitem{Hagen22a}
G.~Hagen, S.~Novario, Z.~Sun, T.~Papenbrock, G.~Jansen, J.~Lietz, T.~Duguet, A.~Tichai, {Angular-momentum projection in coupled-cluster theory: Structure of $^{34}$Mg}, Physical Review C 105~(6) (2022) 064311.

\bibitem{Frosini22c}
M.~Frosini, T.~Duguet, J.-P. Ebran, B.~Bally, H.~Hergert, T.~R. Rodr\'\i{}guez, R.~Roth, J.~Yao, V.~Som\`a, {Multi-reference many-body perturbation theory for nuclei: III. Ab initio calculations at second order in PGCM-PT}, Eur. Phys. J. A 58~(4) (2022) 64.
\newblock \href {http://arxiv.org/abs/2111.01461} {\path{arXiv:2111.01461}}, \href {https://doi.org/10.1140/epja/s10050-022-00694-x} {\path{doi:10.1140/epja/s10050-022-00694-x}}.

\bibitem{Tichai:2023epe}
A.~Tichai, P.~Demol, T.~Duguet, {Towards heavy-mass ab initio nuclear structure: Open-shell Ca, Ni and Sn isotopes from Bogoliubov coupled-cluster theory}, Phys. Lett. B 851 (2024) 138571.
\newblock \href {http://arxiv.org/abs/2307.15619} {\path{arXiv:2307.15619}}, \href {https://doi.org/10.1016/j.physletb.2024.138571} {\path{doi:10.1016/j.physletb.2024.138571}}.

\bibitem{Ro09}
R.~Roth, {Importance truncation for large-scale configuration interaction approaches}, {Physical Review C} {79}~({6}) ({2009}) {064324}.

\bibitem{Tichai:2018qge}
A.~Tichai, J.~M\"uller, K.~Vobig, R.~Roth, {Natural orbitals for ab initio no-core shell model calculations}, Phys. Rev. C 99~(3) (2019) 034321.
\newblock \href {http://arxiv.org/abs/1809.07571} {\path{arXiv:1809.07571}}, \href {https://doi.org/10.1103/PhysRevC.99.034321} {\path{doi:10.1103/PhysRevC.99.034321}}.

\bibitem{Tichai:2019ksh}
A.~Tichai, J.~Ripoche, T.~Duguet, {Pre-processing the nuclear many-body problem: Importance truncation versus tensor factorization techniques}, Eur. Phys. J. A 55~(6) (2019) 90.
\newblock \href {http://arxiv.org/abs/1902.09043} {\path{arXiv:1902.09043}}, \href {https://doi.org/10.1140/epja/i2019-12758-6} {\path{doi:10.1140/epja/i2019-12758-6}}.

\bibitem{Hoppe:2020elo}
J.~Hoppe, A.~Tichai, M.~Heinz, K.~Hebeler, A.~Schwenk, {Natural orbitals for many-body expansion methods}, Phys. Rev. C 103~(1) (2021) 014321.
\newblock \href {http://arxiv.org/abs/2009.04701} {\path{arXiv:2009.04701}}, \href {https://doi.org/10.1103/PhysRevC.103.014321} {\path{doi:10.1103/PhysRevC.103.014321}}.

\bibitem{Porro:2021rjp}
A.~Porro, V.~Som\`a, A.~Tichai, T.~Duguet, {Importance truncation in non-perturbative many-body techniques: Gorkov self-consistent Green\textquoteright{}s function calculations}, Eur. Phys. J. A 57~(10) (2021) 297.
\newblock \href {http://arxiv.org/abs/2103.14544} {\path{arXiv:2103.14544}}, \href {https://doi.org/10.1140/epja/s10050-021-00606-5} {\path{doi:10.1140/epja/s10050-021-00606-5}}.

\bibitem{Hoppe:2021xqv}
J.~Hoppe, A.~Tichai, M.~Heinz, K.~Hebeler, A.~Schwenk, {Importance truncation for the in-medium similarity renormalization group}, Phys. Rev. C 105~(3) (2022) 034324.
\newblock \href {http://arxiv.org/abs/2110.09390} {\path{arXiv:2110.09390}}, \href {https://doi.org/10.1103/PhysRevC.105.034324} {\path{doi:10.1103/PhysRevC.105.034324}}.

\bibitem{Tichai:2018eem}
A.~Tichai, R.~Schutski, G.~E. Scuseria, T.~Duguet, {Tensor-decomposition techniques for ab initio nuclear structure calculations. From chiral nuclear potentials to ground-state energies}, Phys. Rev. C 99~(3) (2019) 034320.
\newblock \href {http://arxiv.org/abs/1810.08419} {\path{arXiv:1810.08419}}, \href {https://doi.org/10.1103/PhysRevC.99.034320} {\path{doi:10.1103/PhysRevC.99.034320}}.

\bibitem{Tichai:2021rtv}
A.~Tichai, P.~Arthuis, K.~Hebeler, M.~Heinz, J.~Hoppe, A.~Schwenk, {Low-rank matrix decompositions for ab initio nuclear structure}, Phys. Lett. B 821 (2021) 136623.
\newblock \href {http://arxiv.org/abs/2105.03935} {\path{arXiv:2105.03935}}, \href {https://doi.org/10.1016/j.physletb.2021.136623} {\path{doi:10.1016/j.physletb.2021.136623}}.

\bibitem{Tichai:2022mqn}
A.~Tichai, P.~Arthuis, K.~Hebeler, M.~Heinz, J.~Hoppe, A.~Schwenk, L.~Zurek, {Least-square approach for singular value decompositions of scattering problems}, Phys. Rev. C 106~(2) (2022) 024320.
\newblock \href {http://arxiv.org/abs/2205.10087} {\path{arXiv:2205.10087}}, \href {https://doi.org/10.1103/PhysRevC.106.024320} {\path{doi:10.1103/PhysRevC.106.024320}}.

\bibitem{Tichai:2023dda}
A.~Tichai, P.~Arthuis, K.~Hebeler, M.~Heinz, J.~Hoppe, T.~Miyagi, A.~Schwenk, L.~Zurek, {Low-Rank Decompositions of Three-Nucleon Forces via Randomized Projections} (7 2023).
\newblock \href {http://arxiv.org/abs/2307.15572} {\path{arXiv:2307.15572}}.

\bibitem{Zare:2022cdw}
A.~Zare, R.~Wirth, C.~A. Haselby, H.~Hergert, M.~Iwen, {Modewise Johnson\textendash{}Lindenstrauss embeddings for nuclear many-body theory}, Eur. Phys. J. A 59~(5) (2023) 95.
\newblock \href {http://arxiv.org/abs/2211.01305} {\path{arXiv:2211.01305}}, \href {https://doi.org/10.1140/epja/s10050-023-00999-5} {\path{doi:10.1140/epja/s10050-023-00999-5}}.

\bibitem{signoracci2015ab}
A.~Signoracci, T.~Duguet, G.~Hagen, G.~Jansen, {Ab initio Bogoliubov coupled cluster theory for open-shell nuclei}, {Physical Review C} {91}~({6}) ({2015}) {064320}.

\bibitem{Hohenstein_2022}
E.~G. Hohenstein, B.~S. Fales, R.~M. Parrish, T.~J. Martínez, \href{http://dx.doi.org/10.1063/5.0077770}{Rank-reduced coupled-cluster. iii. tensor hypercontraction of the doubles amplitudes}, The Journal of Chemical Physics 156~(5) (Feb. 2022).
\newblock \href {https://doi.org/10.1063/5.0077770} {\path{doi:10.1063/5.0077770}}.
\newline\urlprefix\url{http://dx.doi.org/10.1063/5.0077770}

\bibitem{Tichai:2020jpk}
A.~Tichai, R.~Wirth, J.~Ripoche, T.~Duguet, {Symmetry reduction of tensor networks in many-body theory I. Automated symbolic evaluation of $SU(2)$ algebra}, Eur. Phys. J. A 56~(10) (2020) 272.
\newblock \href {http://arxiv.org/abs/2002.05011} {\path{arXiv:2002.05011}}, \href {https://doi.org/10.1140/epja/s10050-020-00233-6} {\path{doi:10.1140/epja/s10050-020-00233-6}}.

\bibitem{Miyagi2021}
T.~Miyagi, S.~R. Stroberg, P.~Navr\'atil, K.~Hebeler, J.~D. Holt, Converged ab initio calculations of heavy nuclei, Phys. Rev. C 105 (2022) 014302.
\newblock \href {https://doi.org/10.1103/PhysRevC.105.014302} {\path{doi:10.1103/PhysRevC.105.014302}}.

\bibitem{RoBi12}
R.~Roth, S.~Binder, K.~Vobig, A.~Calci, J.~Langhammer, P.~Navr\'atil, {Ab Initio Calculations of Medium-Mass Nuclei with Normal-Ordered Chiral NN+3N Interactions}, Physical Review Letters 109 (2012) 052501.

\bibitem{Gebrerufael:2015yig}
E.~Gebrerufael, A.~Calci, R.~Roth, {Open-shell nuclei and excited states from multireference normal-ordered Hamiltonians}, Phys. Rev. C 93~(3) (2016) 031301.
\newblock \href {https://doi.org/10.1103/PhysRevC.93.031301} {\path{doi:10.1103/PhysRevC.93.031301}}.

\bibitem{Frosini21a}
M.~Frosini, T.~Duguet, B.~Bally, Y.~Beaujeault-Taudi\`ere, J.~P. Ebran, V.~Som\`a, {In-medium $k$-body reduction of $n$-body operators: A flexible symmetry-conserving approach based on the sole one-body density matrix}, Eur. Phys. J. A 57~(4) (2021) 151.
\newblock \href {http://arxiv.org/abs/2102.10120} {\path{arXiv:2102.10120}}, \href {https://doi.org/10.1140/epja/s10050-021-00458-z} {\path{doi:10.1140/epja/s10050-021-00458-z}}.

\bibitem{haftel70a}
M.~I. Haftel, F.~Tabakin, {Nuclear saturation and the smoothness of nucleon-nucleon potentials}, Nucl. Phys. A 158 (1970) 1.

\bibitem{Duguet:2003yi}
T.~Duguet, {Bare versus effective pairing forces: A Microscopic finite range interaction for HFB calculations in coordinate space}, Phys. Rev. C 69 (2004) 054317.
\newblock \href {http://arxiv.org/abs/nucl-th/0311065} {\path{arXiv:nucl-th/0311065}}, \href {https://doi.org/10.1103/PhysRevC.69.054317} {\path{doi:10.1103/PhysRevC.69.054317}}.

\bibitem{Robledo:2010ef}
L.~M. Robledo, {Separable approximation to two-body matrix elements}, Phys. Rev. C 81 (2010) 044312.
\newblock \href {http://arxiv.org/abs/1003.4788} {\path{arXiv:1003.4788}}, \href {https://doi.org/10.1103/PhysRevC.81.044312} {\path{doi:10.1103/PhysRevC.81.044312}}.

\bibitem{Frosini22a}
M.~Frosini, T.~Duguet, J.-P. Ebran, V.~Som\`a, {Multi-reference many-body perturbation theory for nuclei: I. Novel PGCM-PT formalism}, Eur. Phys. J. A 58~(4) (2022) 62.
\newblock \href {http://arxiv.org/abs/2110.15737} {\path{arXiv:2110.15737}}, \href {https://doi.org/10.1140/epja/s10050-022-00692-z} {\path{doi:10.1140/epja/s10050-022-00692-z}}.

\bibitem{Ti20}
A.~Tichai, R.~Roth, T.~Duguet, {Many-body perturbation theories for finite nuclei}, {Frontiers in Physics} {8} ({2020}) {164}.

\bibitem{bloch58a}
C.~Bloch, {Sur la d\'etermination de l'\'etat fondamental d'un syst\`eme de particules}, Nucl. Phys. A 7 (1958) 451.

\bibitem{Shavitt2009}
I.~Shavitt, R.~Bartlett, \href{{https://doi.org/10.1017/cbo9780511596834}}{{Many{\textendash}Body Methods in Chemistry and Physics}}, {Cambridge University Press}, {2009}.
\newblock \href {https://doi.org/{10.1017/cbo9780511596834}} {\path{doi:{10.1017/cbo9780511596834}}}.
\newline\urlprefix\url{{https://doi.org/10.1017/cbo9780511596834}}

\bibitem{SiDu15}
A.~Signoracci, T.~Duguet, G.~Hagen, G.~Jansen, {Ab initio Bogoliubov coupled cluster theory for open-shell nuclei}, {Physical Review C} {91}~({6}) ({2015}) {064320}.

\bibitem{Schu17}
R.~Schutski, J.~Zhao, T.~M. Henderson, G.~E. Scuseria, \href{https://doi.org/10.1063/1.4996988}{Tensor-structured coupled cluster theory}, The Journal of Chemical Physics 147~(18) (2017) 184113.
\newblock \href {http://arxiv.org/abs/https://doi.org/10.1063/1.4996988} {\path{arXiv:https://doi.org/10.1063/1.4996988}}, \href {https://doi.org/10.1063/1.4996988} {\path{doi:10.1063/1.4996988}}.
\newline\urlprefix\url{https://doi.org/10.1063/1.4996988}

\bibitem{Duguet:2015yle}
T.~Duguet, A.~Signoracci, {Symmetry broken and restored coupled-cluster theory. II. Global gauge symmetry and particle number}, J. Phys. G 44~(1) (2017) 015103, [Erratum: J.Phys.G 44, 049601 (2017)].
\newblock \href {http://arxiv.org/abs/1512.02878} {\path{arXiv:1512.02878}}, \href {https://doi.org/10.1088/0954-3899/44/1/015103} {\path{doi:10.1088/0954-3899/44/1/015103}}.

\bibitem{Arthuis:2018yoo}
P.~Arthuis, T.~Duguet, A.~Tichai, R.-D. Lasseri, J.-P. Ebran, {ADG: Automated generation and evaluation of many-body diagrams I. Bogoliubov many-body perturbation theory}, {Computer Physics Communications} {240} ({2019}) {202--227}.

\bibitem{Ripoche2020}
J.~Ripoche, A.~Tichai, T.~Duguet, \href{{http://dx.doi.org/10.1140/epja/s10050-020-00045-8}}{{Normal-ordered k-body approximation in particle-number-breaking theories}}, {Eur. Phys. J. A} {56}~({2}) ({2020}).
\newblock \href {https://doi.org/{10.1140/epja/s10050-020-00045-8}} {\path{doi:{10.1140/epja/s10050-020-00045-8}}}.
\newline\urlprefix\url{{http://dx.doi.org/10.1140/epja/s10050-020-00045-8}}

\bibitem{Duguet:2020hdm}
T.~Duguet, B.~Bally, A.~Tichai, {Zero-pairing limit of Hartree-Fock-Bogoliubov reference states}, Phys. Rev. C 102~(5) (2020) 054320.
\newblock \href {http://arxiv.org/abs/2006.02871} {\path{arXiv:2006.02871}}, \href {https://doi.org/10.1103/PhysRevC.102.054320} {\path{doi:10.1103/PhysRevC.102.054320}}.

\bibitem{Braess05}
D.~Braess, W.~Hackbusch, \href{https://doi.org/10.1093/imanum/dri015}{{Approximation of 1/x by exponential sums in [1, infty)}}, IMA Journal of Numerical Analysis 25~(4) (2005) 685--697.
\newblock \href {http://arxiv.org/abs/https://academic.oup.com/imajna/article-pdf/25/4/685/2037151/dri015.pdf} {\path{arXiv:https://academic.oup.com/imajna/article-pdf/25/4/685/2037151/dri015.pdf}}, \href {https://doi.org/10.1093/imanum/dri015} {\path{doi:10.1093/imanum/dri015}}.
\newline\urlprefix\url{https://doi.org/10.1093/imanum/dri015}

\bibitem{tichai2019tf}
A.~Tichai, R.~Schutski, G.~E. Scuseria, T.~Duguet, \href{http://dx.doi.org/10.1103/PhysRevC.99.034320}{Tensor-decomposition techniques for ab initio nuclear structure calculations: From chiral nuclear potentials to ground-state energies}, Physical Review C 99~(3) (Mar. 2019).
\newblock \href {https://doi.org/10.1103/physrevc.99.034320} {\path{doi:10.1103/physrevc.99.034320}}.
\newline\urlprefix\url{http://dx.doi.org/10.1103/PhysRevC.99.034320}

\bibitem{tichai2019pre}
A.~Tichai, J.~Ripoche, T.~Duguet, {Pre-processing the nuclear many-body problem}, {The European Physical Journal A} {55}~({6}) ({2019}) {90}.

\bibitem{martinsson2021randomized}
P.-G. Martinsson, J.~Tropp, Randomized numerical linear algebra: Foundations \& algorithms (2021).
\newblock \href {http://arxiv.org/abs/2002.01387} {\path{arXiv:2002.01387}}.

\bibitem{tropp2023randomized}
J.~A. Tropp, R.~J. Webber, Randomized algorithms for low-rank matrix approximation: Design, analysis, and applications (2023).
\newblock \href {http://arxiv.org/abs/2306.12418} {\path{arXiv:2306.12418}}.

\bibitem{Nakatsukasa07a}
T.~Nakatsukasa, T.~Inakura, K.~Yabana, {Finite amplitude method for the solution of the random-phase approximation}, {Physical Review C} {76}~({2}) ({2007}) {024318}.

\bibitem{Carlsson_2012}
B.~G. Carlsson, J.~Toivanen, A.~Pastore, \href{http://dx.doi.org/10.1103/PhysRevC.86.014307}{Collective vibrational states within the fast iterative quasiparticle random-phase approximation method}, Physical Review C 86~(1) (Jul. 2012).
\newblock \href {https://doi.org/10.1103/physrevc.86.014307} {\path{doi:10.1103/physrevc.86.014307}}.
\newline\urlprefix\url{http://dx.doi.org/10.1103/PhysRevC.86.014307}

\bibitem{Hebeler11a}
K.~Hebeler, S.~K. Bogner, R.~J. Furnstahl, A.~Nogga, A.~Schwenk, {Improved nuclear matter calculations from chiral low-momentum interactions}, Phys. Rev. C 83 (2011) 031301.
\newblock \href {http://arxiv.org/abs/1012.3381} {\path{arXiv:1012.3381}}, \href {https://doi.org/10.1103/PhysRevC.83.031301} {\path{doi:10.1103/PhysRevC.83.031301}}.

\bibitem{Miyagi2023}
T.~Miyagi, \href{http://dx.doi.org/10.1140/epja/s10050-023-01039-y}{Nuhamil: A numerical code to generate nuclear two- and three-body matrix elements from chiral effective field theory}, The European Physical Journal A 59~(7) (Jul. 2023).
\newblock \href {https://doi.org/10.1140/epja/s10050-023-01039-y} {\path{doi:10.1140/epja/s10050-023-01039-y}}.
\newline\urlprefix\url{http://dx.doi.org/10.1140/epja/s10050-023-01039-y}

\bibitem{Scalesi24}
S.~et. al, in preparation (2024).

\bibitem{SoDu11}
V.~Som{\`a}, T.~Duguet, C.~Barbieri, {Ab initio self-consistent Gorkov-Green's function calculations of semimagic nuclei: Formalism at second order with a two-nucleon interaction}, {Physical Review C} {84}~({6}) ({2011}) {064317}.

\bibitem{tichai2021adg}
A.~Tichai, P.~Arthuis, H.~Hergert, T.~Duguet, {ADG: automated generation and evaluation of many-body diagrams: III. Bogoliubov in-medium similarity renormalization group formalism}, The European Physical Journal A 58~(1) (2022) 2.

\end{thebibliography}

\end{document}